\documentclass[12pt]{article}

\usepackage{ametsoc}
\usepackage[english]{babel}
\usepackage[latin1]{inputenc}
\usepackage{times}
\usepackage[T1]{fontenc}
\usepackage{pgf}

\usepackage{hyperref}
\hypersetup{colorlinks=true,urlcolor=blue,linkcolor=black}

\def\labelPrefix{GridStd}

\def\missref#1{\texttt{(missing ref: #1)}} 
\newcommand{\balaji}{ V. Balaji\footnote{  
    Corresponding author: V. Balaji,
    Princeton University and NOAA/GFDL,
    201 Forrestal Road, Princeton NJ 08540-6646.
    Email: \texttt{balaji@princeton.edu}} \\
  Princeton University
}

\newcommand{\degree}{\ensuremath{^\circ}}
\newcommand{\codeBlock}[2]{     
  \begin{equation}
    \texttt{\framebox{\shortstack[l]{#2}}}{\label{code:#1}}
  \end{equation}
}

\renewcommand\today{\number\day\space \ifcase\month\or
  January\or February\or March\or April\or May\or June\or
  July\or August\or September\or October\or November\or December\fi
  \space\number\year}

\newcommand{\bibref}[1] {\cite{ref:#1}}
\ifx\labelPrefix\undefined
\newcommand{\ceqref}[1] {\mbox{CodeBlock \ref{code:#1}}}

\newcommand{\figref}[1] {\mbox{Figure  \ref{fig:#1}}}
\newcommand{\secref}[1] {\mbox{Section \ref{sec:#1}}}
\newcommand{\tabref}[1] {\mbox{Table   \ref{tab:#1}}}
\else
\newcommand{\ceqref}[1] {\mbox{CodeBlock \ref{code:\labelPrefix:#1}}}

\newcommand{\figref}[1] {\mbox{Figure  \ref{fig:\labelPrefix:#1}}}
\newcommand{\secref}[1] {\mbox{Section \ref{sec:\labelPrefix:#1}}}
\newcommand{\tabref}[1] {\mbox{Table   \ref{tab:\labelPrefix:#1}}}
\fi

\usepackage[strings]{underscore}         

\title{Gridspec: A standard for the description of grids used in Earth
  System models}
\author{
  \balaji \and
  Alistair Adcroft \\
  Princeton University \\ \and
  Zhi Liang \\
  NOAA/Geophysical Dynamics Laboratory
}
\date\today

\begin{document}

\maketitle

\abstract{
  The comparative analysis of output from multiple models, and against
  observational data analysis archives, has become a key methodology
  in reducing uncertainty in climate projections, and in improving
  forecast skill of medium- and long-term forecasts. There is
  considerable momentum toward simplifying such analyses by applying
  comprehensive community-standard metadata to observational and model
  output data archives.
  
  The representation of gridded data is a critical element in
  describing the contents of model output. We seek here to propose a
  standard for describing the grids on which such data are
  discretized. The standard is drafted specifically for inclusion
  within the Climate and Forecasting (CF) metadata conventions.
  
    The contents of this paper have been in the ``grey literature''
  since 2007: it has been posted to arXiv to be citable. To preserve
  its integrity, the contents have not been updated.
}

\tableofcontents

\section{Introduction}
\label{sec:GridStd:Intro}

\subsection{Methodology of international modeling campaigns}
\label{sec:GridStd:Intro:Methodology}

The current decade (2000-2010) may be regarded as the decade of the
coming-of-age of Earth System models. Such models are coming into
routine use in both research and operational settings: for
understanding the planetary climate in terms of feedbacks and balances
between its many components; for translating such understanding into
projections that inform policy to address anthropogenic climate
change; and increasingly for medium- and long-term forecasts that
require coupled models as well.

These activities manifest themselves in aspects of current scientific
methodology. Earth System science is becoming ``big science'' where
experiments systematically involve large international modeling
campaigns, matching in scale the observational campaigns that are
responsible for producing the climate record. A key example of such a
modeling campaign is the activity surrounding the Inter-Governmental
Panel on Climate Change (IPCC) Assessment Reports. These reports,
issued every 6 years, are a culmination of systematic and coordinated
modeling experiments run at multiple institutions around the world.
\figref{IPCC} shows a list of participating IPCC institutions from the
recently concluded Fourth Assessment Report (IPCC AR4) \missref{}. A
comparative study of results from multiple models run under the same
external forcings remains our best tool for understanding the climate
system, and for generating consensus and uncertainty estimates of
climate change. Several key papers based on the IPCC AR4 data archive
at PCMDI document recent leaps in understanding of aspects of the
climate system in stable and warming climates, such as ENSO
\citep{ref:guilyardietal2009,ref:oldenborghetal2001}, the tropical
circulation \citep[e.g][]{ref:vecchietal2006}, Southern ocean
circulation \citep{ref:russelletal2006}, and others.
Other similar campaigns underway include the Aqua-Planet Experiment
(APE) \missref{}, the ENSEMBLES project \citep{ref:hewittgriggs2004}
as well as several older ones.

\begin{figure}[htb]
  \centering
  \includegraphics*[width=140mm]{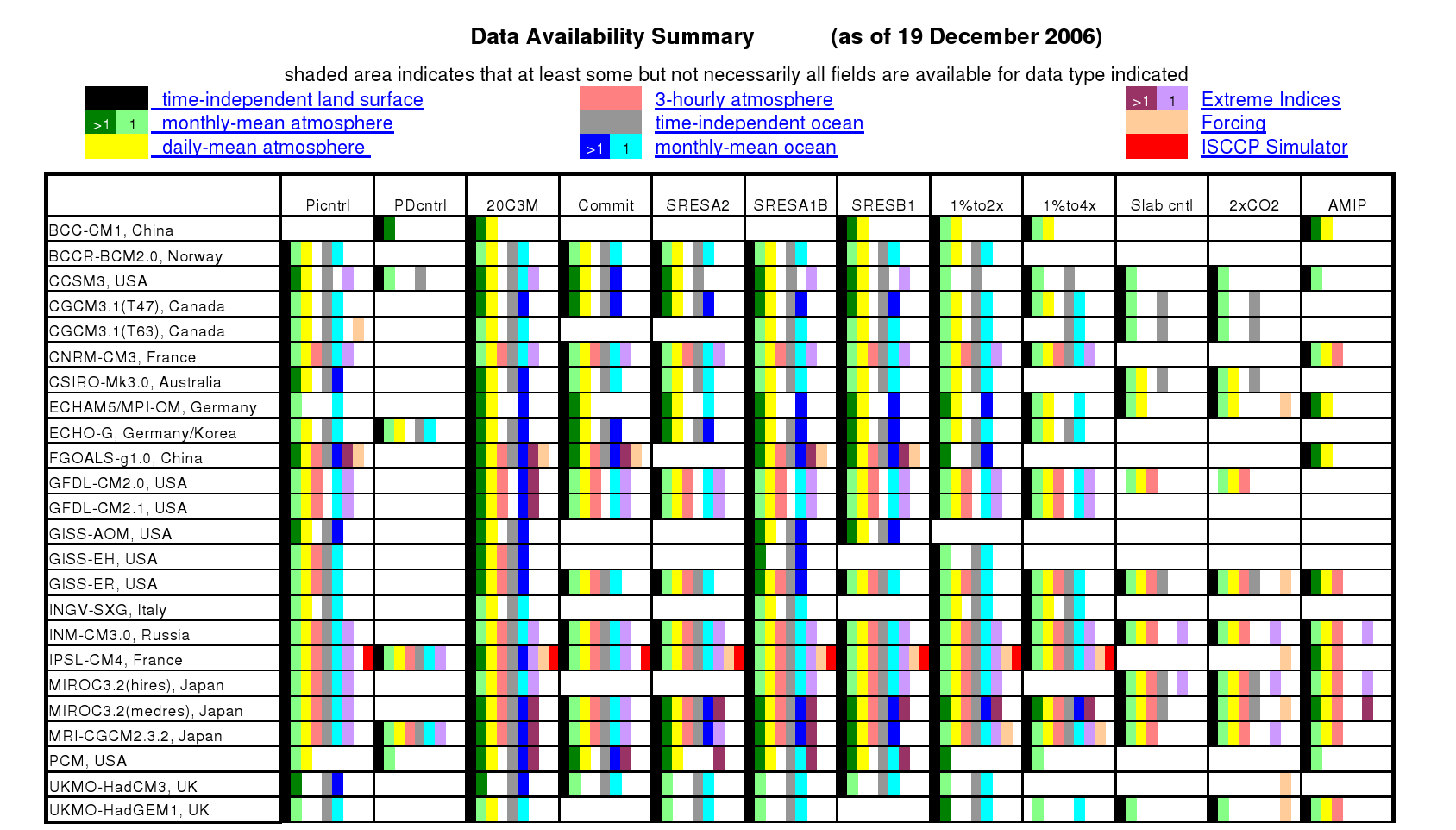}
  \caption{Participating institutions in the IPCC AR4 series of experiments.}
  \label{fig:GridStd:IPCC}
\end{figure}

It has also become apparent that a similar multi-model ensemble
approach is of utility in seasonal and interannual forecasting as
well. An example of such a modeling campaign is the DEMETER project
\citep{ref:palmeretal2004}. Studies
\citep[e.g][]{ref:hagedornetal2005} show that such operational ensemble
forecasts have demonstrably better forecast skill than any individual
ensemble member.

A third trend in current modeling studies is the increased use of
\emph{downscaling}, reviewed in \bibref{wilbywigley1997}. Where
fine-scale simulation over some domain is sought, and it is
either useless (because there is limited impact of fine-scale structure
on larger scales) or impractical (for computational reasons) to extend
the high resolution over the entire domain, one often creates model
chains, where models over larger domains at coarser resolution are
used to force finer-scale models nested within. The use of model
chains is also a sort of multi-model study, where output data from one
model serves as input to another. In all the approaches above, the
need for data standards to enable ready access to data from diverse
models is apparent.

\subsection{Community approaches to models and data}
\label{sec:GridStd:Community}

As Earth System science increasingly comes to depend on models created
from multiple components, and on comparative studies of output from
such models, standardization has become a serious issue as we grapple
with the practicalities of carrying out such studies. Emerging efforts
at standardization of model component interfaces include the Earth
System Modeling Framework (ESMF) \citep[ESMF:
][]{ref:hilletal2004,ref:collinsetal2005}.

Model output data in the Earth System Science community increasingly
converges on the
\href{http://www.unidata.ucar.edu/software/netcdf}{netCDF format},
and, to a lesser degree, the \href{http://hdf.ncsa.uiuc.edu/HDF5}{HDF5
  format}. In the weather forecasting domain, the WMO-mandated GRIB
and BUFR formats \missref{} continue to be used. While the data
formats themselves are relatively mature, recent efforts in this
domain focus on developing consistent and comprehensive
\emph{metadata}, data descriptors that provide human- and
machine-readable information about the data necessary in interpreting
its contents. Metadata vocabularies are intended eventually to enable
the inclusion of data into a \emph{semantic web}
\citep{ref:berners-lee1999,ref:berners-leeetal2001,ref:berners-leehendler2001}
which human and other reasoning agents will be able to use to make
useful inferences about found entities. In the climate and weather
modeling domain, efforts at developing a common vocabulary for
metadata have converged on the Climate and Forecasting (CF)
conventions. Similar initiatives for observational data (e.g the
\href{http://marinemetadata.org}{Marine Metadata Initiative (MMI)})
abound, and there are attempts underway to align the CF vocabularies
with the observational ones. The
\href{http://www.opengeospatial.org}{Open Geospatial Consortium (OGC)}
is a possible mechanism to shepherd the CF conventions toward a formal
standard.

\subsection{Rationale for a grid standard}
\label{sec:GridStd:Intro:Rationale}

This paper focuses on a key element of the metadata under development:
the \emph{grids} on which model data is discretized. Experience from
the international modeling campaigns cited above in
\secref{Intro:Methodology} indicates that there is a wide diversity in
the model grids used; and further, it appears that this diversity is
only increasing. However, in the absence of a standard representation
of grids, it has been rather difficult to perform comparative analyses
of data from disparate model grids. Rather, the lead institutions in
these campaigns insist upon having data delivered on very simple
grids, on the credible argument that the sites running the models are
best placed to perform regridding operations of appropriate quality,
meeting the relevant scientific criteria of conservation, and so on.

This approach was followed in the IPCC AR4 campaign, and while the
resulting data archive was an extraordinary boon to data
\emph{consumers} (analysts of model output), the burden it placed on
data \emph{producers} (modeling centres) was considerable. Further, the
issues surrounding regridding are common to most modeling centres,
capable of being abstracted to common software. We believe a suite of
common regridding methods and tools is now possible, given a grid
standard.

The grid standard becomes even more necessary in considering the other
sorts of uses outlined in \secref{Intro:Methodology}, such as in model
chains where gridded data from one model becomes input to another. And
last but not least, multiple model grids and data transformations
between them are intrinsic to modern Earth System models themselves,
and are the basis for coupled model development from components
developed across the entire community.

This paper proposes a \emph{grid standard}: a convention for
describing model grids. We have described so far its general features
and purposes:

\begin{itemize}
\item the standard will describe the grids commonly used in Earth
  system models from global scale to fine scale, and also with an eye
  looking forward (toward emerging discrete representations) and
  sideways (to allied research domains: space weather, geosciences);
\item the standard will contain all the information required to enable
  commonly performed scientific analysis and visualization of data;
\item the standard will contain all the information required to
  perform transformations from one model grid to another, satisfying
  constraints of conservation and preservation of essential features,
  as science demands;
\item the standard will make possible the development of shared
  regridding software, varying from tools deployable as web services
  to perform on-the-fly regridding from data archives, to routines to
  be used within coupled models. It will enable, but not mandate, the
  use of these standard techniques.
\end{itemize}

An outline of such a grid standard is the topic of this paper.

\subsection{Overview of paper}
\label{sec:Intro:Overview}

The paper is structured as follows. In \secref{Grids} we survey the
types of grids currently in use, and potentially to be used in
emerging models, that the standard must cover. This includes the issue
of vector fields and staggered grids. In \secref{Mosaic} we develop
the key abstractions of mosaics, required for handling nested grids
and other ``non-standard'' tilings of the sphere. In \secref{XGrid} we
cover the issue of masks and exchange grids, required for
transformations of data between grids. In \secref{CF} we develop a
vocabulary for describing grids in the context of the CF conventions.

\section{Grid terminology for Earth System science}
\label{sec:GridStd:Grids}

We begin by developing a terminology for describing the types of grids
used in Earth System science models and datasets. Grids for Earth
System science can be considerably specialized with respect to the
more general grids used in computational fluid dynamics. Specifically,
the vertical extent is considerably smaller ($\sim$10 km) than the
horizontal ($\sim$1000 km), and the fluid in general strongly
stratified in the vertical. The treatment of the vertical is thus
generally separable; and model grids can generally be described
separately in terms of a horizontal 2D grid with coordinates $X$ and
$Y$, and a vertical coordinate $Z$.

\subsection{Geometry}
\label{sec:GridStd:Geometry}

The underlying \emph{geometry} being modeled is most often a thin
spherical shell\footnote{Except at very fine scales, the geometry is
  treated as a sphere, not a geoid. This may be a problem when
  geo-referencing to very precise datasets that consider the surface
  as a geoid.}, especially when it is the actual planetary
dynamics that is being modeled. However, more idealized studies may
use geometries that simplify the rotational properties of the fluid,
such as an $f$-plane or $\beta$-plane, or even simply a cartesian
geometry.

Where the actual Earth or planetary system is being modeled,
\emph{geospatial mapping} or \emph{geo-referencing} is used to map
model coordinates to standard spatial coordinates, usually
\emph{geographic longitude} and \emph{latitude}. Vertical mapping to
pre-defined levels (e.g height, depth or pressure) is also often employed
as a standardization technique when comparing model outputs to each
other, or to observations.

\subsection{Vertical coordinate}
\label{sec:GridStd:VertCoord}

The vertical coordinate can be \emph{space-based} (height or depth
with respect to a reference surface) or \emph{mass-based} (pressure,
density, potential temperature). \emph{Hybrid} coordinates with a
mass-based element are considered to be mass-based.

The \emph{reference surface} is a digital elevation map of the
planetary surface. This can be a detailed topography or bathymetry
digital elevation dataset, or a more idealized one such as the
representation of a single simplified mountain or ridge, or none at
all. Vertical coordinates requiring a reference surface are referred
to as \emph{terrain-following}. Both space-based (\citep[e.g
Gal-Chen,SLEVE][]{ref:gal-chensomerville1975,ref:schaeretal2002} and
mass-based (e.g $\sigma$) terrain-following coordinates are commonly
used.

The rationale for developing this minimal taxonomy to classify vertical
coordinates is that translating one class of vertical coordinate into
another is generally model- and problem-specific, and should
\emph{not} be attempted by standard regridding software.

\subsection{Horizontal coordinates}
\label{sec:GridStd:HorizCoord}

Horizontal spatial coordinates may be \emph{polar} ($\theta$,$\phi$)
coordinates on the sphere, or \emph{planar} ($x$,$y$), where the
underlying geometry is cartesian, or based on one of several
\emph{projections} of a sphere onto a plane. Planar coordinates based
on a spherical projection define a \emph{map factor} allowing a
translation of ($x$,$y$) to ($\theta$,$\phi$).

\emph{Curvilinear coordinates} may be used in both the polar and
planar instances, where the model refers to a pseudo-longitude and
latitude, that is then mapped to geographic longitude and latitude by
geo-referencing. Examples include the displaced-pole grid
\citep{ref:jonesetal2005} and the tripolar grid
\citep{ref:murray1996}.

Horizontal coordinates may have the important properties of
\emph{orthogonality} (when the $Y$ coordinate is normal to the $X$)
and \emph{uniformity} (when grid lines in either direction are
uniformly spaced). Numerically generated grids may not be able to
satisfy both constraints simultaneously.

A third type of horizontal coordinate often used in this domain is not
spatial, but spectral. \emph{Spectral coordinates} on the sphere
represent the horizontal distribution of a variable in terms of its
spherical harmonic coefficients. These coefficients can be uniquely
mapped back and forth to polar coordinates based on Fourier and
Legendre transforms, yielding uniformly spaced longitudes, and
latitudes defined by a Gaussian quadrature. This grid specification
will not consider spectral representations directly; rather, it
assumes that the data have been transformed to polar coordinates, and
only seeks to encode the \emph{truncation} used to restrict the
representation to a finite set of values.

Spectral coordinates on the plane have also recently been used in this
domain. These methods generally employ \emph{spectral elements}
\citep{ref:thomasloft2002,ref:iskandaranietal2002} projecting
the sphere onto a series of planes of finite spatial extent, within
each of which the representation is spectral. Spectral elements are
also uniquely bound to geospatial coordinates by a series of
transforms, and it is in these coordinates that the data are assumed
to have been written.

\subsection{Time coordinate}
\label{sec:GridStd:Time}

As for the fourth coordinate, time, it is already reasonably
well-covered in the CF conventions. Both instantaneous and
time-averaged values are represented. Key issues that still remain
include the definition and treatment of non-standard \emph{calendars},
and for simulation data, a standard vocabulary to define aspects of a
running experiment, such as the absolute start time of the simulation.

\subsection{Discretization}
\label{sec:GridStd:Discretization}

In translating a data variable to a discrete representation, we must
decide what aspects are necessary for inclusion in a standard grid
specification. We have chosen two classes of operations that the grid
standard must enable: \emph{vector calculus}, differential and
integral operations on scalar and vector fields; and
\emph{conservative regridding}, the transformation of a variable from
one grid to another in a manner that preserves chosen moments of its
distribution, such as area and volume integrals of 2D and 3D scalar
fields. We recognize that higher-order methods that preserve variances
or gradients may entail some loss of accuracy. In the case of vector
fields, grid transformations that preserve streamlines are required.

To enable vector calculus and conservative regridding, the following
aspects of a grid must be included in the specification:

\begin{itemize}
\item \emph{distances} between gridpoints, to allow differential operations;
\item \emph{angles} of grid lines with respect to a reference, usually
  geographic East and North, to enable vector operations. One may also
  choose to include an \emph{arc type} (e.g ``great circle''), which
  specifies families of curves to follow while integrating a grid line
  along a surface.
\item \emph{areas} and \emph{volumes} for integral operations. This is
  generally done by defining the boundaries of a grid cell represented
  by a point value. In \secref{XGrid} below we will also consider
  fractional areas and volumes in the presence of a \emph{mask}, which
  defines the sharing of cell between two or more components.
\end{itemize}

A taxonomy of grids may now be defined. A discretization is
\emph{logically rectangular} if the coordinate space $(x,y,z)$ is
translated one-to-one to index space \texttt{(i,j,k)}. Note that the
coordinate space may continue to be physically curvilinear; yet, in
index space, grid cells will be rectilinear boxes.

The most commonly used discretization in Earth system science is
logically rectangular, and that will remain the principal object of
study here. Beyond the simplest logically rectangular grids may
include more specialized grids such as the tripolar grid of
\bibref{murray1996} shown in \figref{tripolar} and the cubed-sphere
grid of \bibref{rancicetal1996}, shown in \figref{cube}.

\begin{figure}[htb]
  \centering
  \includegraphics*[height=64mm]{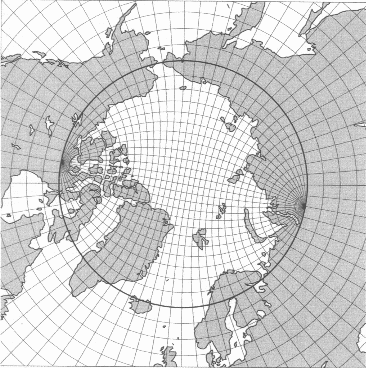}
  \caption{The tripolar grid, often used in ocean modeling. Polar
    singularities are placed over land and excluded from the
    simulation.}
  \label{fig:GridStd:tripolar}
\end{figure}

\begin{figure}[htb]
  \centering
  \includegraphics*[height=64mm]{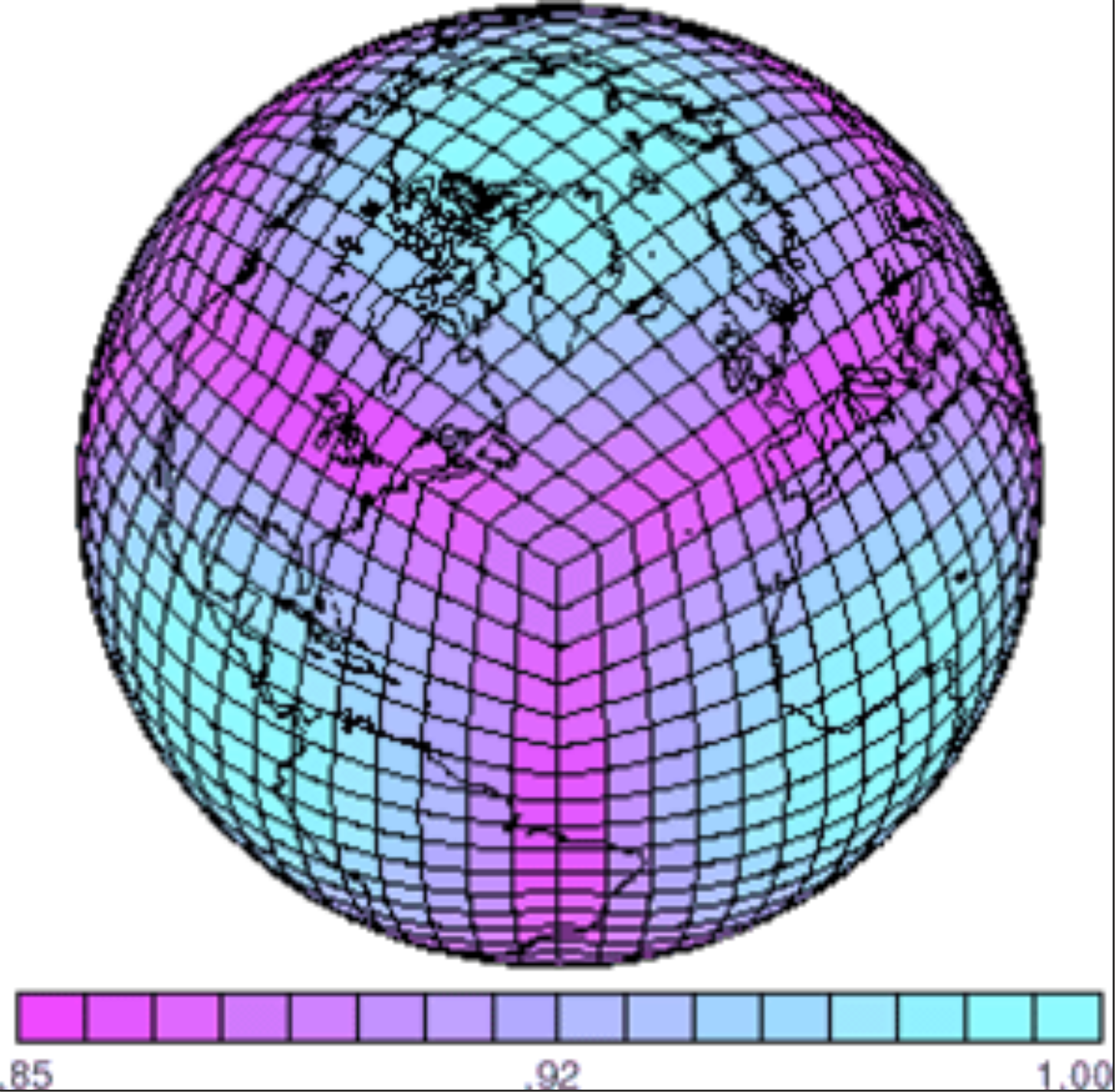}
  \caption{The cube-sphere grid, projecting the sphere onto the six
    faces of a cube. Polar singularities are avoided, at the expense
    of some grid distortion near the cube's vertices.}
  \label{fig:GridStd:cube}
\end{figure}

Triangular discretizations are increasingly voguish in the field. A
\emph{structured triangular} discretization of an icosahedral
projection is a popular new approach resulting in a geodesic grid
\citep{ref:majewskietal2002,ref:randalletal2002}. An example of a
structured triangular grid is shown in \figref{iko} from
\bibref{majewskietal2002}. The grid is generated by recursive division
of the 20 triangular faces of an icosahedron.

\begin{figure}[htb]
  \centering
  \includegraphics*[height=64mm]{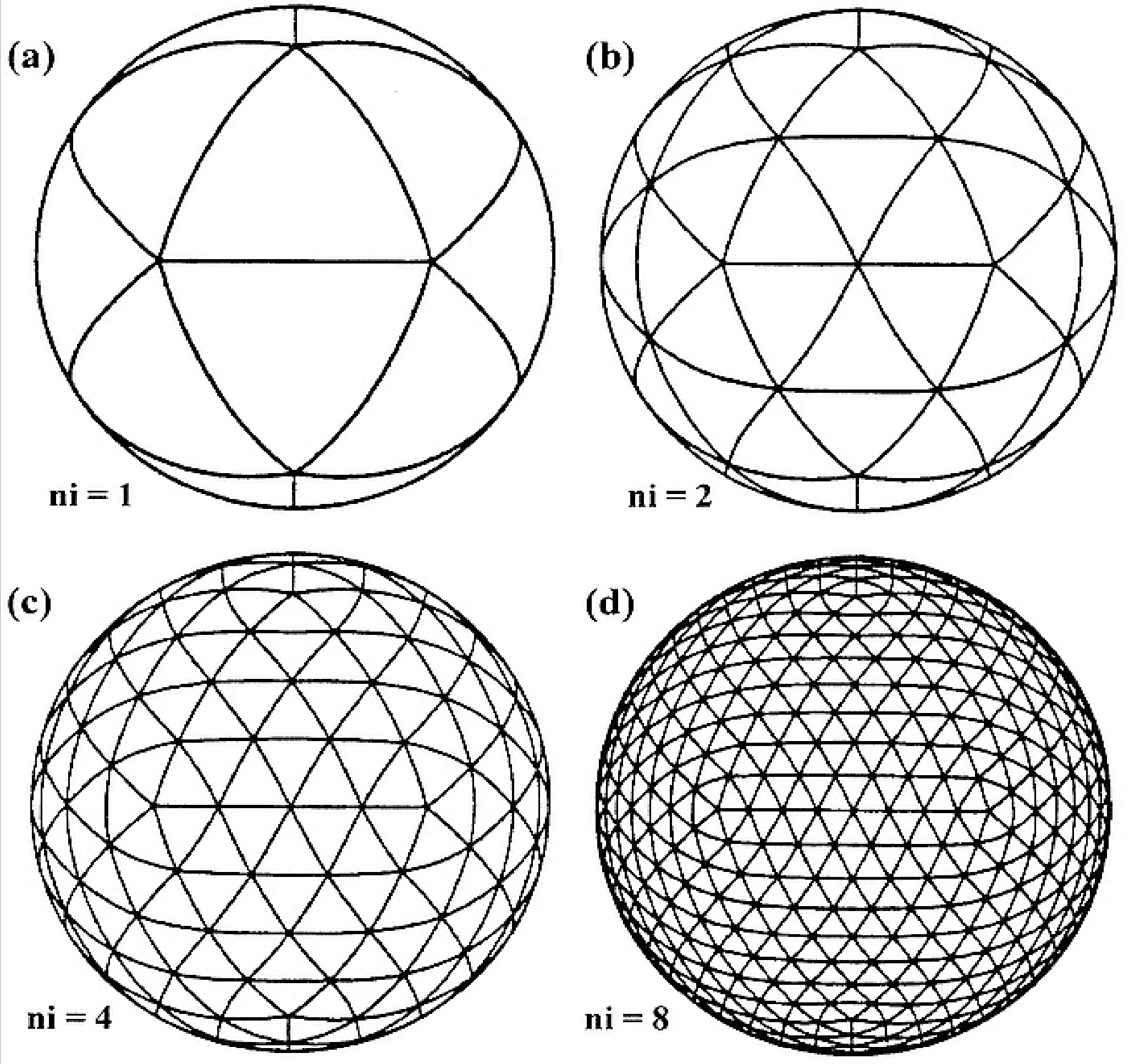}
  \caption{A structured triangular discretization of the sphere. Note
    that all vertices at any truncation level \texttt{ni} are also
    vertices at any higher level of truncation.}
  \label{fig:GridStd:iko}
\end{figure}

Numerically generated \emph{unstructured triangular} discretization,
such as shown in \figref{utgExample} are often used, especially over
complex terrain. High resolution models interacting with real
topography increasingly use such unstructured grids.
\secref{Unstructured} visits the issue of the specification of such
grids.

\begin{figure}[htb]
  \centering
  \includegraphics*[width=152mm]{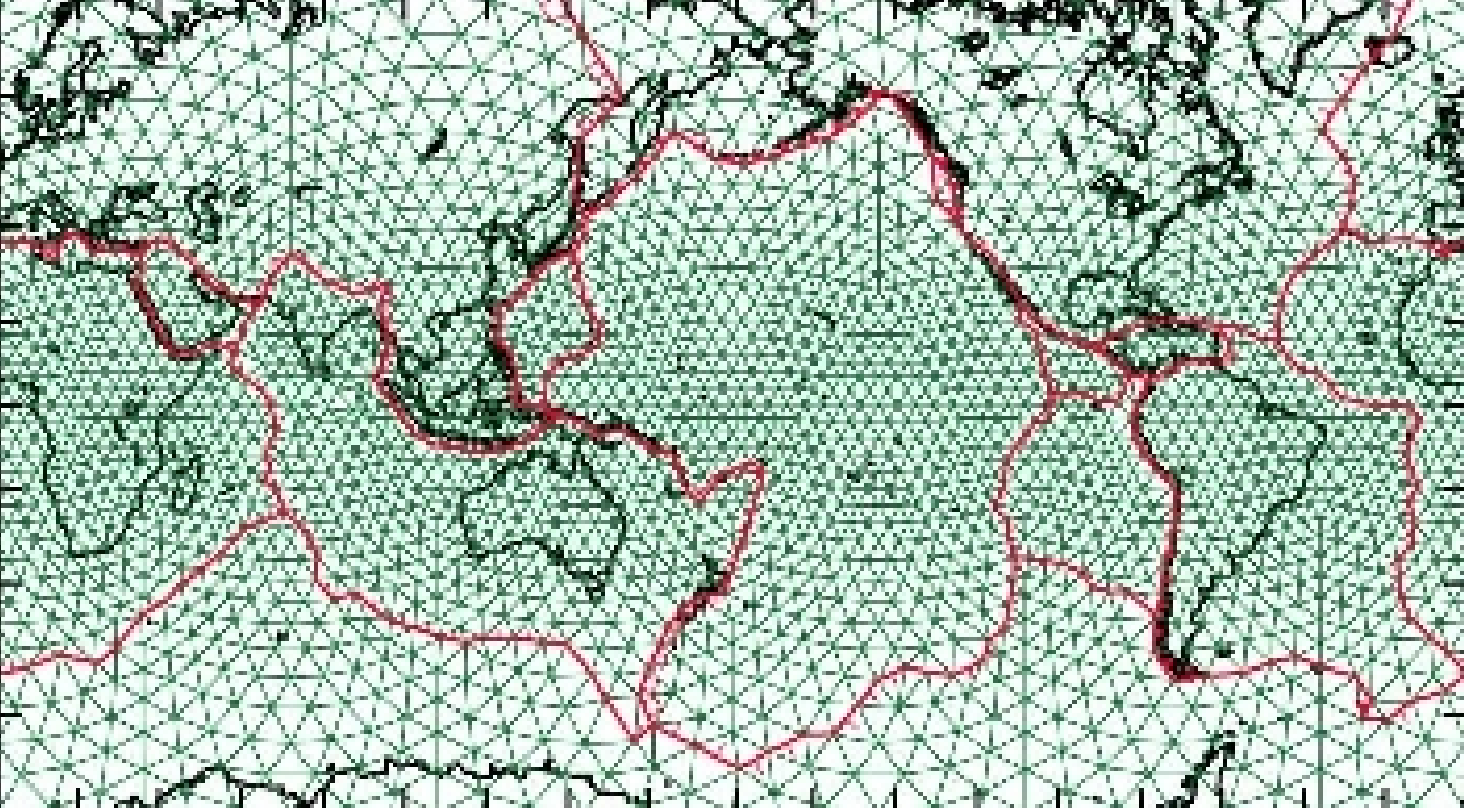}
  \caption{An unstructured triangular discretization of the sphere.}
  \label{fig:GridStd:utgExample}
\end{figure}

There is no need for unstructured grids to have only triangular
elements (although we shall see in \secref{Staggering} that the
\emph{supergrid} abstraction allows us to build all such grids out of
UTGs). \emph{Unstructured polygonal grids} of arbitrary polygonal
elements are a completely general abstraction, where each cell might
have any number of vertices. In practice, we usually find somewhat
more restrictive formulations such as in
\href{http://oceanmodeling.rsmas.miami.edu/seom/seom_intro.html}{Spectral
  Element Ocean Model (SEOM)} of \bibref{iskandaranietal2002} cited
earlier: an example SEOM grid for the ocean is shown in \figref{seom}.

\begin{figure}[htb]
  \centering
  \includegraphics*[width=120mm]{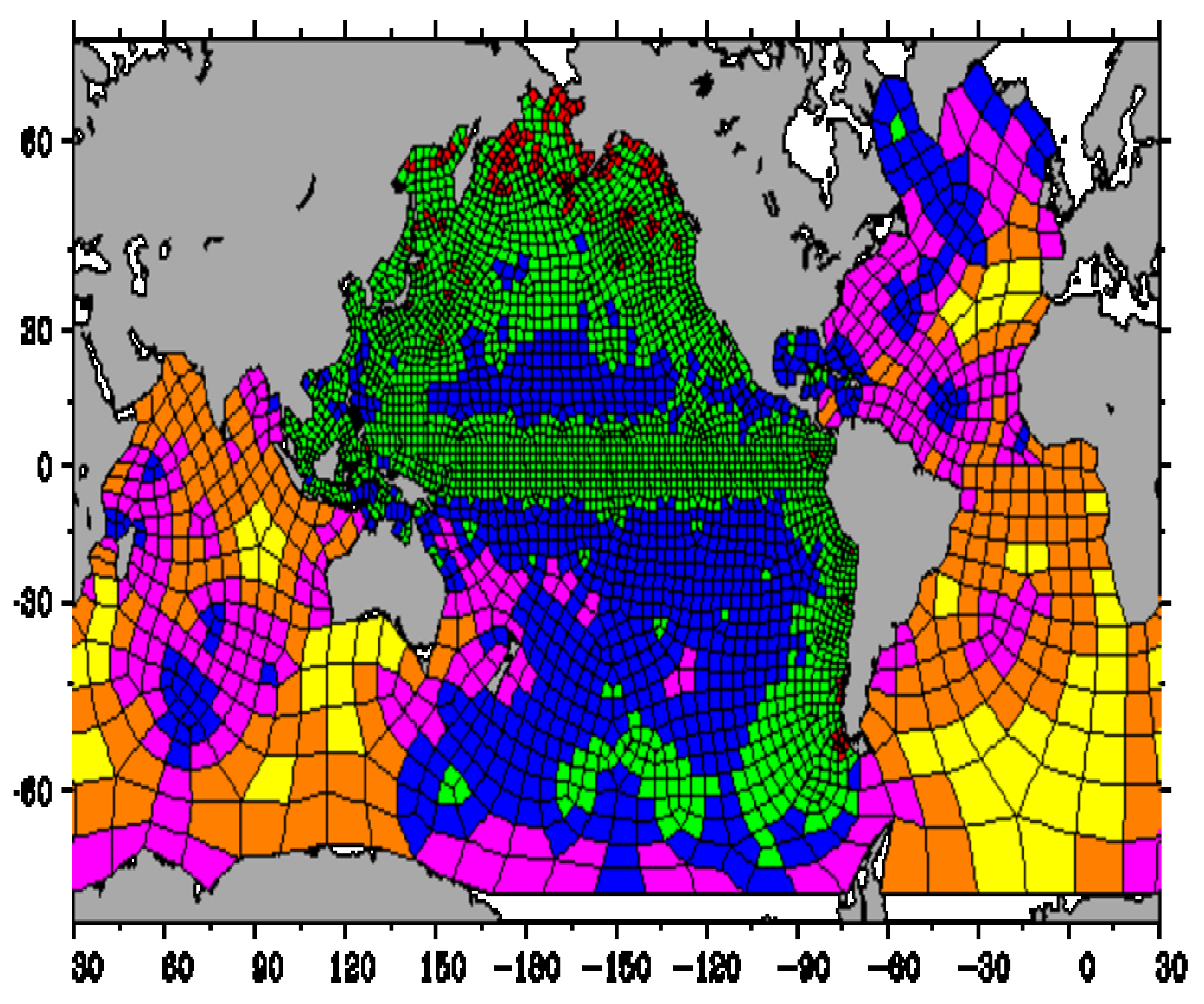}
  \caption{The SEOM unstructured grid.}
  \label{fig:GridStd:seom}
\end{figure}

A reasonably complete taxonomy of grid discretizations for the near-
to mid-future in Earth System science would include:

\begin{description}
\item[LRG] logically rectangular grid.
\item[STG] structured triangular grid.
\item[UTG] unstructured triangular grid.
\item[UPG] unstructured polygonal grid.
\item[PCG] pixel-based catchment grids: gridboxes made up of arbitrary
  collections of contiguous fine-grained pixels, usually used to
  demarcate \emph{catchments} defined by surface elevation isolines
  \citep{ref:kosteretal2000}.
\item[EGG] Escher gecko grid.
\end{description}

\begin{figure}[htb]
  \centering
  \includegraphics*[width=100mm]{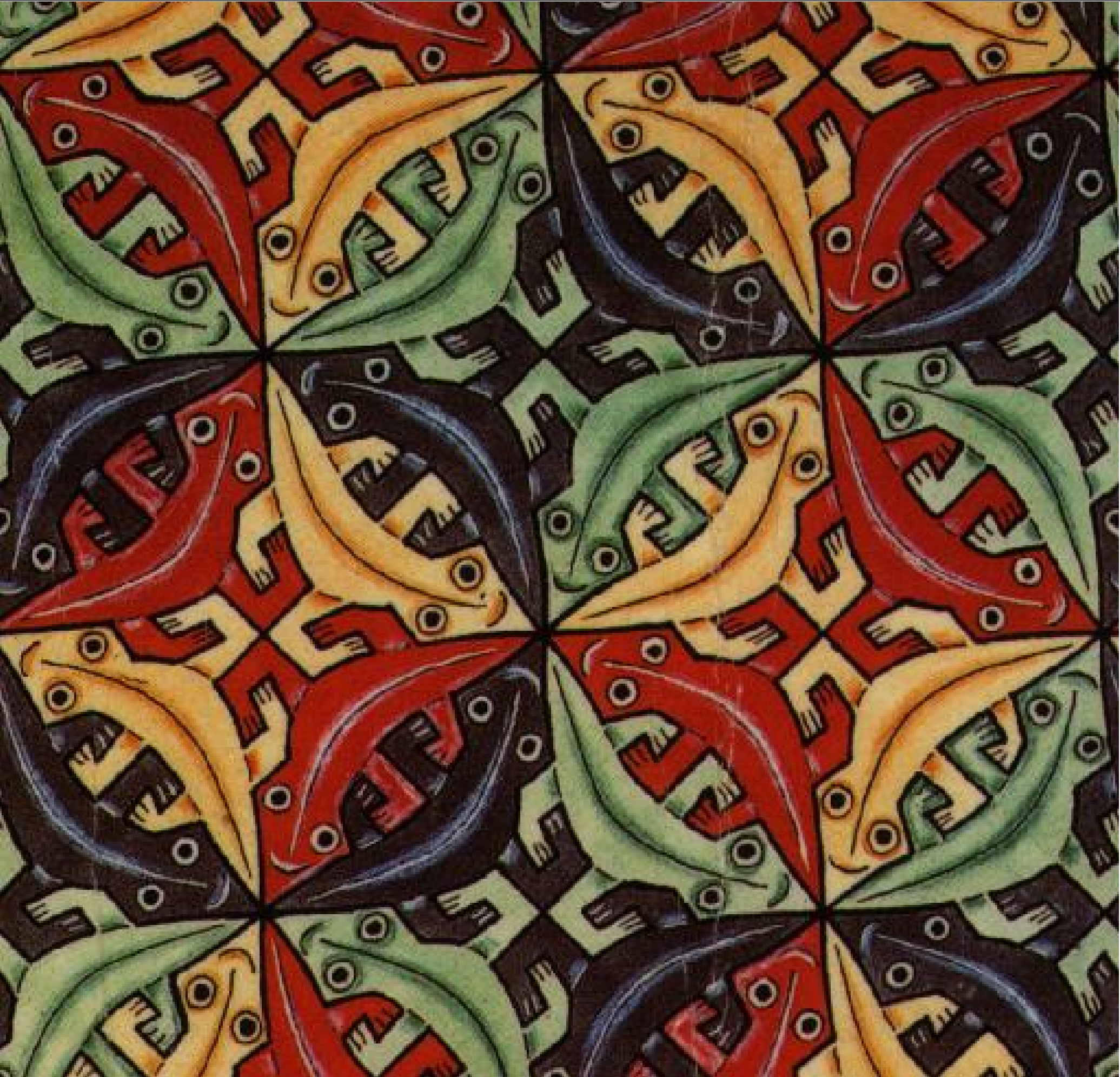}
  \caption{Another possible discretization of the plane.}
  \label{fig:GridStd:eschergecko}
\end{figure}

While developing a vocabulary and placeholders for all of the above,
we shall focus here principally on logically rectangular
discretizations. We shall expose the key concepts of \emph{supergrids}
(\secref{Staggering}) and \emph{mosaics} (\secref{Mosaic}) based on
LRGs, and aim to show their relevance for other discretization types
as well. We expect the specification to be extended to these other
discretization types by the relevant domain experts, as in
\secref{Unstructured}.

\subsection{Staggering, refinement, and the supergrid}
\label{sec:GridStd:Staggering}

Algorithms place quantities at different locations within a grid
cell (``staggering''). In particular, the Arakawa grids, covered in
standard texts such as \bibref{haltinerwilliams1980} show different
ways to represent velocities and masses on grids, as shown in
\figref{stagger}.

\begin{figure}[htb]
  \setlength{\unitlength}{0.7cm}
  \centering
  \begin{picture}(23,6){
      \multiput(0,1)(6,0){4}{
        \put(0,0){\framebox(4,4){}}
        \put(2,2){\circle*{0.25}}
        \put(2,2){\makebox(0.6,0.6){\textbf{T}}}
      }
      \put(2,3){\thicklines\color{red}\vector(1,0){0.8}}
      \put(2.6,3){\makebox(0.6,0.6){\textbf{U}}}
      \put(2,3){\thicklines\color{red}\vector(0,1){0.8}}
      \put(2,3.6){\makebox(0.6,0.6){\textbf{V}}}
      \multiput(6,1)(4,0){2}{
        \multiput(0,0)(0,4){2}{
          \put(0,0){\circle*{0.25}}
          \put(0,0){\thicklines\color{red}\vector(0,1){0.8}}
          \put(0.6,0){\makebox(0.6,0.6){\textbf{U}}}
          \put(0,0){\thicklines\color{red}\vector(1,0){0.8}}
          \put(0,0.6){\makebox(0.6,0.6){\textbf{V}}}
        }
      }
      \multiput(12,3)(4,0){2}{
        \put(0,0){\circle*{0.25}}
        \put(0,0){\thicklines\color{red}\vector(1,0){0.8}}
        \put(0.6,0){\makebox(0.6,0.6){\textbf{U}}}
      }
      \multiput(14,1)(0,4){2}{
        \put(0,0){\circle*{0.25}}
        \put(0,0){\thicklines\color{red}\vector(0,1){0.8}}
        \put(0,0.6){\makebox(0.6,0.6){\textbf{V}}}
      }
      \multiput(18,3)(4,0){2}{
        \put(0,0){\circle*{0.25}}
        \put(0,0){\thicklines\color{red}\vector(0,1){0.8}}
        \put(0,0.6){\makebox(0.6,0.6){\textbf{V}}}
      }
      \multiput(20,1)(0,4){2}{
        \put(0,0){\circle*{0.25}}
        \put(0,0){\thicklines\color{red}\vector(1,0){0.8}}
        \put(0.6,0){\makebox(0.6,0.6){\textbf{U}}}
      }
      \put( 1.7,0){\makebox(0.6,0.6){\texttt{A}}}
      \put( 7.7,0){\makebox(0.6,0.6){\texttt{B}}}
      \put(13.7,0){\makebox(0.6,0.6){\texttt{C}}}
      \put(19.7,0){\makebox(0.6,0.6){\texttt{D}}}
    }
  \end{picture}
  
  \caption{The Arakawa staggered grids.}
  \label{fig:GridStd:stagger}
\end{figure}
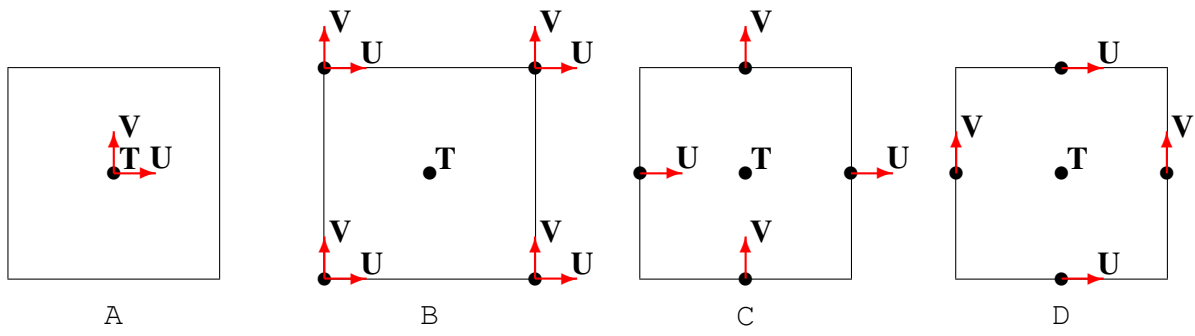

This has led to considerable confusion in terminology and design: are
the velocity and mass grids to be constructed independently, or as
aspects (``subgrids'') of a single grid? How do we encode the
relationships between the subgrids, which are necessarily fixed and
algorithmically essential?

In this approach, we dispense with subgrids, and instead invert
the specification: we define a \emph{supergrid}. The supergrid is an
object potentially of higher refinement than the grid that an
algorithm will use; but every such grid needed by an application is a
subset of the supergrid.

Given a complete specification of distances, angles, areas and volumes
on a supergrid, any operation on any Arakawa grid is completely
defined.

The refinement of an Arakawa grid is always 2: here we generalize the
refinement factor to an arbitrary integer, so that a single
high-resolution grid specification may be used to run simulations at
different resolutions.

We can now define a \emph{cell} without ambiguity: it is an element of
a supergrid. A \emph{cell} on the grid itself may be overspecified,
but this guarantees that any set of staggered grids will have
consistent coordinate distances and areas.

The supergrid cell itself does not have a ``center'': in constructing
a grid from a supergrid, the grid center is indeed a vertex on the
supergrid. However, certain applications of supergrids require the
specification of a \emph{centroid} \citep[e.g][]{ref:jones1999}, a
representative cell location. This is nominally some the center of
some weighting field distributed about its area; but it is incorrect
to try and compute a distance from centroid to a vertex.

Staggered arrays may be defined as \emph{symmetric} or
\emph{asymmetric arrays}. Taking the Arakawa C-grid (\figref{cgrid})
as an example, we have a 8$\times$8 supergrid. Scalars, at cell
centres, will form a 4$\times$4 array. A \emph{symmetric array}
representing the velocity component $U$ will be of size 5$\times$4.
Quite often, though, all arrays may be defined to be 4$\times$4, in
which case, one must also specify if the $U$ points are biased to the
``east'' or ``west'', i.e if the array value \texttt{u(i,j)} refers to
the point $U(i+\frac12,j)$ or $U(i-\frac12,j)$. While this can be
inferred from the array size, it is probably wise to include this
information in the specification for readability.

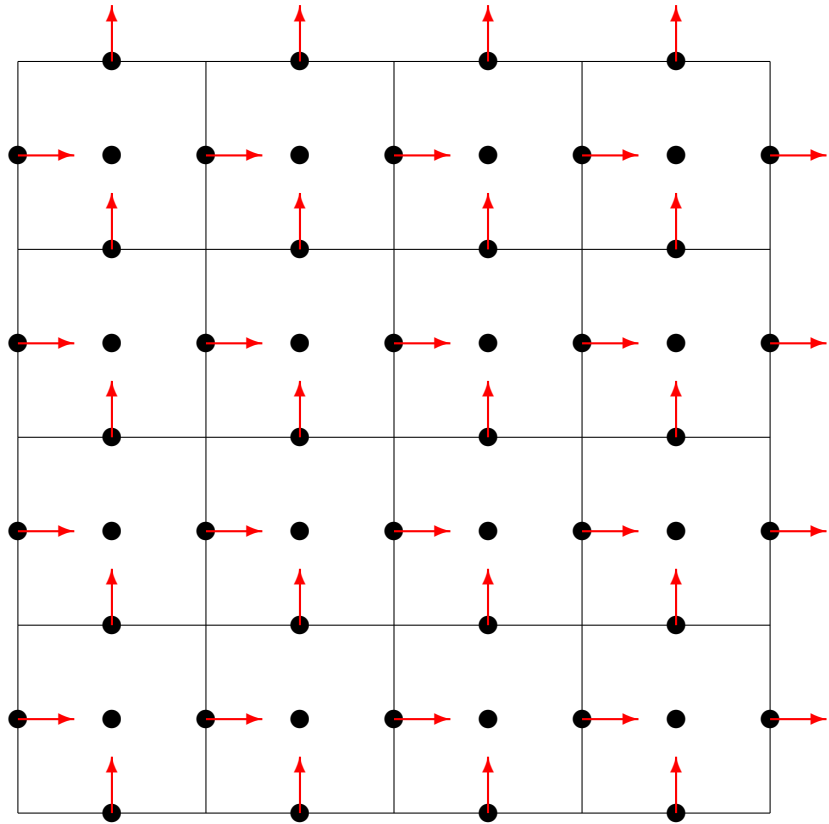
\begin{figure}[htb]
  \setlength\unitlength{2.5cm}
  \centering
  \begin{picture}(4,4.5)
    \multiput(0,0)(1,0){4}{
      \multiput(0,0)(0,1){4}{
        \put(0,0){\line(1,0){1}}
        \put(0,0){\line(0,1){1}}
        \put(0.5,0.5){\circle*{0.1}}
        \put(0,0.5){\circle*{0.1}}
        \put(0,0.5){\thicklines\color{red}\vector(1,0){0.3}}
        \put(0.5,0){\circle*{0.1}}
        \put(0.5,0){\thicklines\color{red}\vector(0,1){0.3}}
      }
    }
    \put(0,4){\line(1,0){4}}
    \put(4,0){\line(0,1){4}}
    \multiput(0,4)(1,0){4}{
      \put(0.5,0){\circle*{0.1}}
      \put(0.5,0){\thicklines\color{red}\vector(0,1){0.3}}
    }
    \multiput(4,0)(0,1){4}{
      \put(0,0.5){\circle*{0.1}}
      \put(0,0.5){\thicklines\color{red}\vector(1,0){0.3}}
    }
  \end{picture}
  
  \caption{A 4$\times$4 (not 8$\times$8!) Arakawa C-grid.}
  \label{fig:GridStd:cgrid}
\end{figure}

Grid \emph{refinement} is another application of supergrids. A refined
grid is usually a fine grid overlying a coarse grid, with some integer
factor of resolution in index space. The vertices on the coarse grid
are also vertices on the fine grid, as shown in the example of
\figref{refined}.

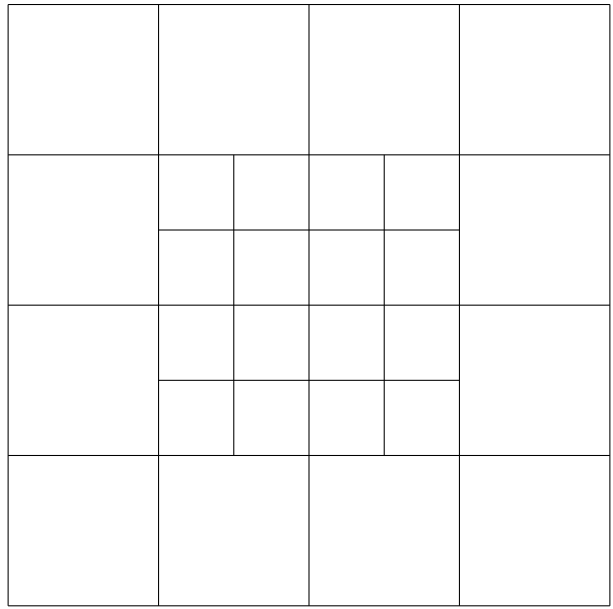
\begin{figure}[htb]
  \setlength{\unitlength}{10mm}
  \centering
  \begin{picture}(8,8)
    \multiput(0,0)(2,0){5}{\line(0,1){8}}
    \multiput(0,0)(0,2){5}{\line(1,0){8}}
    \multiput(3,2)(2,0){2}{\line(0,1){4}}
    \multiput(2,3)(0,2){2}{\line(1,0){4}}
  \end{picture}
  \caption{Nested grids with integer refinement: an inner 4$\times$4
    grid at twice the resolution is nested within the coarse
    4$\times$4 grid.}
  \label{fig:GridStd:refined}
\end{figure}

The coincidence of certain vertices of refined grids in contact permit
certain operations more specialized than the completely generalized
overlap contact region specified in \secref{XGrid}. The supergrid
plays a role here, as the vertices of a single logically rectangular
supergrid can capture all of the grid information for a refined grid.
Of course, adaptive refinement techniques where grids may be
indefinitely refined may not allow for the prior definition of that
supergrid.

\subsubsection{Triangular supergrids}

Can the supergrid idea be extended to non-rectangular grids? It is
somewhat less intuitive in this case, but it is argued in this article
that the supergrid idea is equally applicable to grids that are not
logically rectangular. There are several reasons to attempt to encode
unstructured grids in this fashion. First, we see in the STG of
\figref{iko} that coarse resolution grids, say at \texttt{ni} = 1, 2
or 4, can be constructed by subsampling a supergrid defined at
\texttt{ni} = 8. Second, staggering is a concept equally at home on
triangular grids. It is common practice on STGs and UTGs to define
vertex-, cell-, and face-centered quantities. Furthermore, several key
interpolative algorithms on UTGs depend on these quantities, as shown
in \figref{ikoStagger} from \bibref{majewskietal2002}.

\begin{figure}[htb]
  \centering
  \includegraphics*[width=111mm]{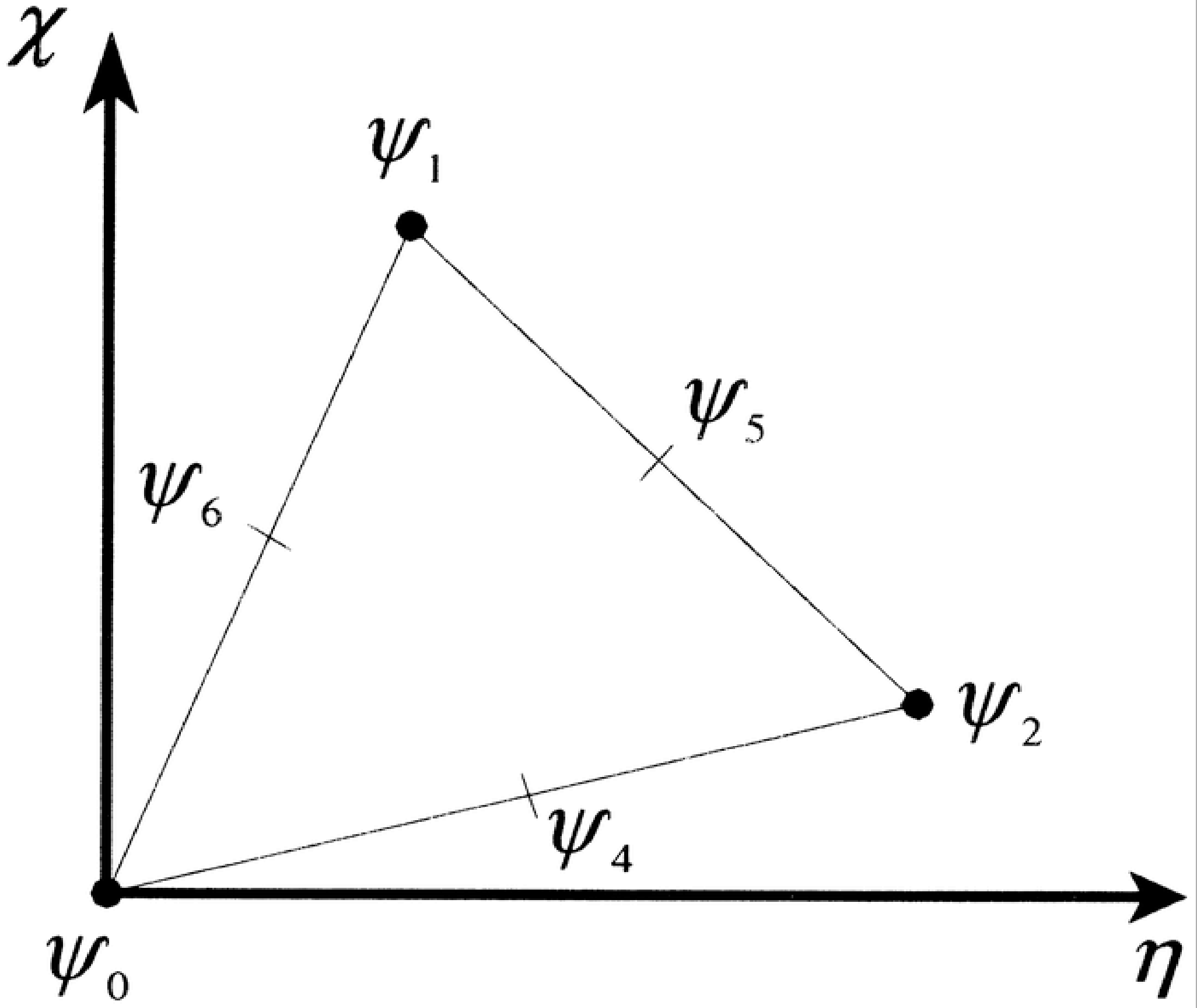}
  \caption{Vertex- and face-centered locations on a triangular grid.
  All of these quantites are needed for certain accurate interpolation
  algorithms on these grids. Further, different quantities may be
  placed on a different subset of points associated with this cell.}
  \label{fig:GridStd:ikoStagger}
\end{figure}

The proposed treatment of unstructured grids, detailed below in
\secref{Unstructured}, is to define a specification of UTGs that
represent a supergrid, i.e including all vertex-, cell-, and
face-centered locations. \emph{Only UTGs need to be considered in
  defining a supergrid, as a triangular supergrid underlies any
  unstructured grid, including those containing polygons with
  arbitrary vertex counts.}

\subsubsection{Raster grids}

Raster grids are a discretization of a surface into high-resolution
pixels of an atomic nature: a ``point'' is the location of its
containing raster, and any ``line'' is made up of discrete segments
that follow raster edges but which cannot intersect them. The
``area'' of any grid cell on a raster is defined merely by counting
the pixels within its bounding curve.

An application of raster grids is the use of catchment grids or PCGs
\citep{ref:kosteretal2000}. Catchment grids follow digital elevation
isolines to form bounding boxes following topography to facilitate
modeling land surface processes. PCGs are defined entirely in terms of
an underlying raster grid.

A raster grid can also be defined on the basis of a high-resolution
supergrid. Typically, these are created on the basis of
high-resolution digital elevation datasets defined on a sphere. Thus
raster grids are defined here as LRG supergrids. The centroid defines
the raster location.

\subsection{Mosaics}
\label{sec:GridStd:Mosaic}

In many applications, it makes sense to divide up the model into a set
of \emph{grid tiles}\footnote{The words \emph{grid} and \emph{tile}
  separately are overused, and can mean many things depending on
  context. We will somewhat verbosely try always to use the term
  \emph{grid tile} to avoid ambiguity.}, each of which is
independently discretized. An example above is the cubed-sphere of
\figref{cube}, which is defined by six grid tiles, on which a data
field may be represented by several arrays, one per tile. We call such
a collection of grid tiles a grid mosaic, as shown in
\figref{Mosaic}.

\begin{figure}[htb]
  \setlength{\unitlength}{0.4cm}
  \centering
  \begin{picture}(20,12)
    \put(0,4){\color{red}\framebox(4,4){}}
    \multiput(1,4)(1,0){3}{\line(0,1){4}}
    \multiput(0,5)(0,1){3}{\line(1,0){4}}
    \multiput(8,0)(4,4){3}{
      \multiput(0,0)(4,0){2}{
        \put(0,0){\color{red}\framebox(4,4){}}
        \multiput(1,0)(1,0){3}{\line(0,1){4}}
        \multiput(0,1)(0,1){3}{\line(1,0){4}}
      }
    }
  \end{picture}

  \caption{A grid tile: a quadrilateral grid shown in index space. A
    grid mosaic: a number of tiles sharing boundaries or contact
    regions.}
  \label{fig:GridStd:Mosaic}
\end{figure}
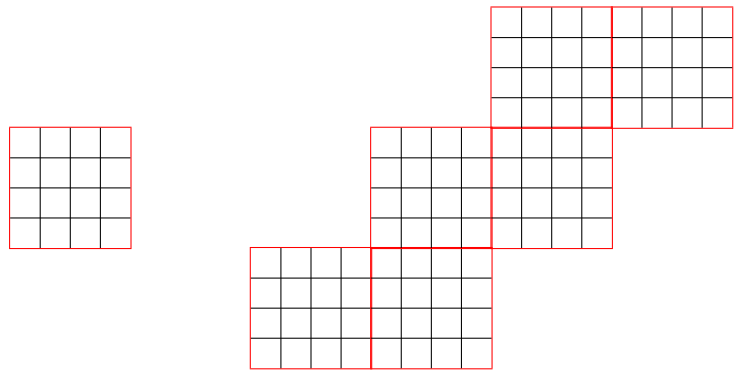

A \emph{grid mosaic} is constructed recursively by referring to child
mosaics, with the tree terminating in leaves defined by \emph{grid
  tiles} (\figref{mosaicTree}).

\begin{figure}[htb]
  \setlength{\unitlength}{10mm}
  \centering
  \newsavebox{\mosaic}
  \savebox{\mosaic}(1,2){
    \thicklines
    \put(0.5,1.5){\circle{1}}
    \put(0,1){\makebox(1,1){M}}
    \put(0.5,0){\line(0,1){1}}
  }
  \newsavebox{\grid}
  \savebox{\grid}(1,1){
    \thicklines
    \put(0,0){\framebox(1,1){G}}
  }
  \begin{picture}(10,8)
    \put(4,6){\usebox{\mosaic}}
    \put(2,6){\line(1,0){6}}
    \multiput(2,5)(3,0){3}{\line(0,1){1}}
    \multiput(1,3)(3,0){3}{\usebox{\mosaic}}
    \multiput(1,2)(6,0){2}{\usebox{\grid}}
    \put(4,3){\line(1,0){2}}
    \multiput(4,2)(2,0){2}{\line(0,1){1}}
    \multiput(3,1)(2,0){2}{\usebox\grid}
  \end{picture}
  \caption{A grid mosaic $M$ is constructed hierarchically; each
    branch of the tree terminates in a grid tile $G$.}
  \label{fig:GridStd:mosaicTree}
\end{figure}
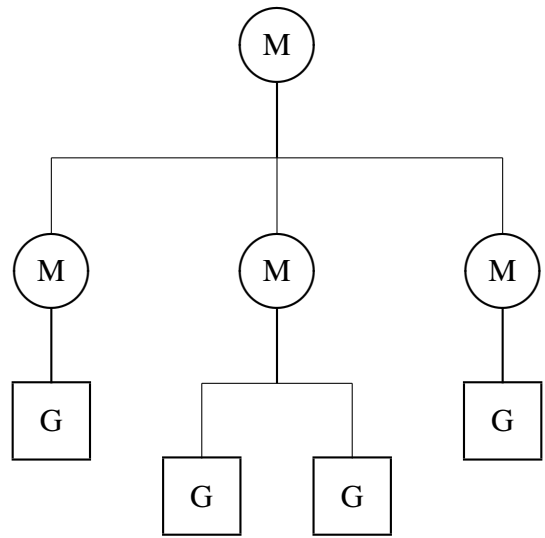

Aside from the grid information in the grid tiles, the grid mosaic
additionally specifies connections between pairs of tiles in the form
of \emph{contact regions} between \emph{pairs} of grid
tiles.\footnote{It is not necessarily possible to deduce contact
  regions by geospatial mapping: there can be applications where
  geographically collocated regions do \emph{not} exchange data, and
  also where there is implicit contact between non-collocated
  regions.}

Contact regions can be \emph{boundaries}, topologically of one
dimension less than the grid tiles (i.e, planes between volumes, or
lines between planes), or \emph{overlaps}, topologically equal in
dimension to the grid tile. In the cubed-sphere example the contact
regions between grid tiles are 1D boundaries: other grids may contain
tiles that overlap. In the example of the \emph{yin-yang} grid
\citep{ref:kageyamaetal2004} of \figref{yinyang} the grid mosaic
contains two grid tiles that are each lon-lat grids, with an overlap.
The overlap is also specified in terms of a \emph{contact region}
between pairs of grid tiles. Issues relating to boundaries are
described in \secref{Boundary}. Overlaps are described in terms of an
exchange grid \citep[e.g][]{ref:balajietal2006}, outlined in
\secref{XGrid}.

\begin{figure}[htb]
  \centering
  \includegraphics*[width=100mm]{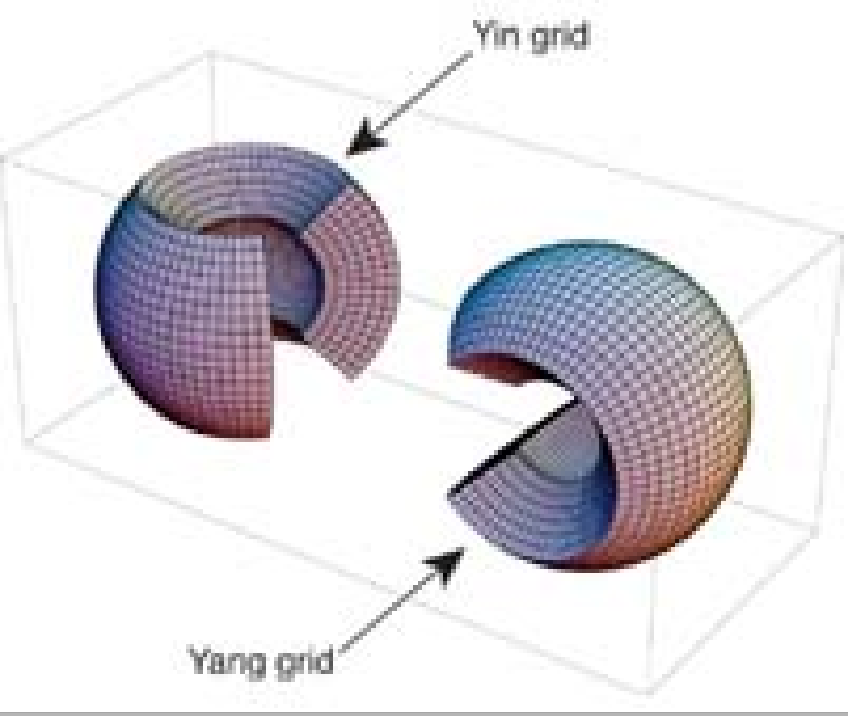}
  \caption{The yin-yang grid consists of two longitude-latitude bands
    with mutually orthogonal axes, and an overlap.}
  \label{fig:GridStd:yinyang}
\end{figure}

The grid mosaic is a powerful abstraction making possible an entire
panoply of applications. These include:

\begin{itemize}
\item the use of overset grids such as the yin-yang grid of
  \figref{yinyang}; 
\item the representation of nested grids \citep[e.g][see
  \figref{cubesphereAMR}]{ref:kuriharaetal1990};
\item the representation of reduced grids
  \citep[e.g][]{ref:rasch1994}. Currently these typically use
  full arrays and a specification of the ``ragged edge''. A reduced
  grid can instead be written as a grid mosaic where each reduction
  appears as a separate grid tile.
\item An entire coupled model application or dataset can be
  constructed as a hierarchical mosaic. Grid mosaics representing
  atmosphere, land, ocean components and so on, as well as contact
  regions between them, all can be represented using this abstraction.
  This approach is already in use at many modeling centres including
  GFDL, though not formalized.
\item Finally, grid mosaics can be used to overcome performance
  bottlenecks associated with parallel I/O and very large files.
  Representing the model grid by a mosaic permits one to save data to
  multiple files, and the step of \emph{aggregation} is deferred. This
  approach is already used at GFDL to perform distributed I/O from a
  parallel application, where I/O aggregation is deferred and
  performed on a separate I/O server sharing a filesystem with the
  compute server.
\end{itemize}

\begin{figure}[htb]
  \centering
  \includegraphics*[height=140mm]{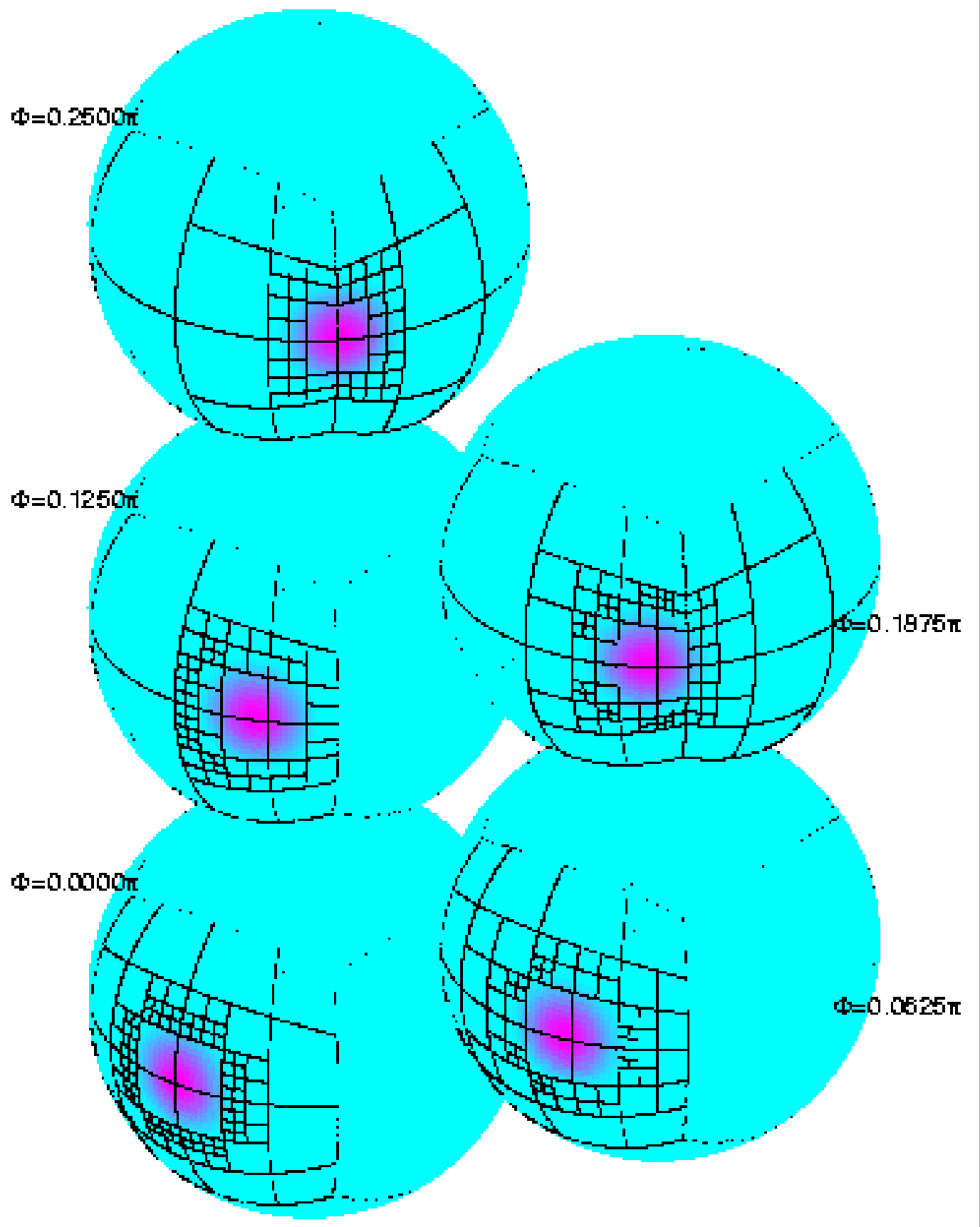}
  \caption{A cubed-sphere with embedded nests.}
  \label{fig:GridStd:cubesphereAMR}
\end{figure}

All of these applications make the grid mosaic abstraction central to
this specification.

\subsection{Boundary contact regions}
\label{sec:GridStd:Boundary}

\emph{Boundaries} for LRG tiles are specified in terms of an
\emph{anchor point} and an \emph{orientation}. An anchor point is a
boundary point that is common to the two grid tiles in contact. When
possible, it is specified as integers giving index space locations of
the anchor point on the two grid tiles. When there is no common grid
point, the anchor point is specified in terms of floating point
numbers giving a geographic location. The \emph{orientation} of the
boundary specifies the index space direction of the running boundary
on each grid tile.

\figref{cubeMosaic} shows an example of boundaries for the
cubed-sphere grid mosaic. Colored lines show shared boundaries between
pairs of grid tiles: note how orientation may change so that a
``north'' edge on one grid tile may be in contact with a ``west'' edge
of another. Orientation changes indicate how vector quantities are
transformed when transiting a grid tile boundary.

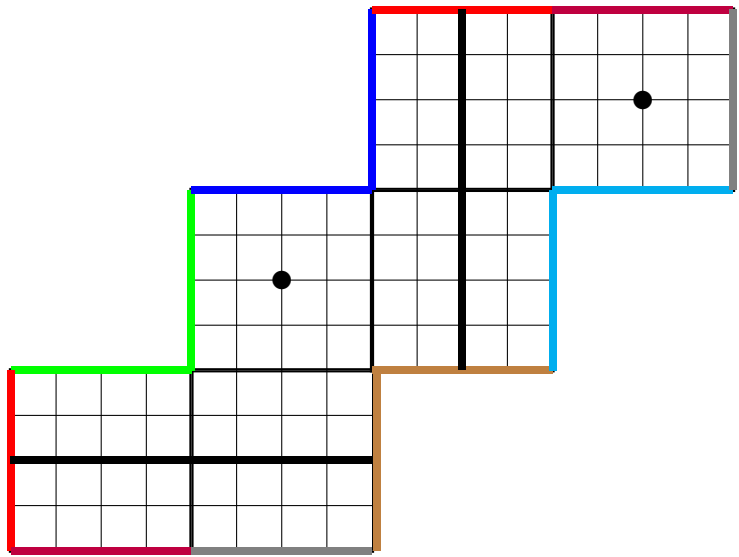
\begin{figure}[htb]
  \setlength\unitlength{0.6cm}
  \centering
  \begin{picture}(16,12)
    \multiput(0,0)(4,0){2}{
      \multiput(0,0)(4,4){3}{
        \put(0,0){\thicklines\framebox(4,4){}}}}
    \multiput(0,1)(4,4){3}{
      \multiput(0,0)(0,1){3}{\line(1,0){8}}}
    \multiput(1,0)(4,4){3}{
      \multiput(0,0)(4,0){2}{
        \multiput(0,0)(1,0){3}{\line(0,1){4}}}}
    {\linethickness{1mm}
      \put(0,4){\color{green}\line(1,0){4}}
      \put(4,4){\color{green}\line(0,1){4}}
      \put(4,8){\color{blue}\line(1,0){4}}
      \put(8,8){\color{blue}\line(0,1){4}}
      \put(0,0){\color{red}\line(0,1){4}}
      \put(8,12){\color{red}\line(1,0){4}}
      \put(8.1,0){\color{brown}\line(0,1){4}}
      \put(8,4){\color{brown}\line(1,0){4}}
      \put(12,4){\color{cyan}\line(0,1){4}}
      \put(12,8){\color{cyan}\line(1,0){4}}
      \put(0,0){\color{purple}\line(1,0){4}}
      \put(12,12){\color{purple}\line(1,0){4}}
      \put(4,0){\color{gray}\line(1,0){4}}
      \put(16,8){\color{gray}\line(0,1){4}}
      \put(0,2){\line(1,0){8}}
      \put(10,4){\line(0,1){8}}
    }
    \multiput(6,6)(8,4){2}{\circle*{0.4}}
  \end{picture}

  \caption{The cubed-sphere grid mosaic.}
  \label{fig:GridStd:cubeMosaic}
\end{figure}

Note that cyclic boundary conditions can be expressed as a contact
region of a grid tile with itself, on opposite edges, and the polar
fold in \figref{tripolar} likewise.

Boundary conditions are considerably simplified when certain
assumptions about grid lines can be made. These are illustrated in
\figref{cubeRefined} for various types of boundaries.

\begin{figure}[htb]
  \setlength\unitlength{0.6cm}
  \centering
  \begin{picture}(16,12)
    \multiput(0,0)(4,4){3}{
      \multiput(0,0)(4,0){2}{
        \put(0,0){\color{red}\framebox(4,4){}}
      }
      \put(4,0){
        \multiput(1,0)(1,0){3}{\line(0,1){4}}
        \multiput(0,1)(0,1){3}{\line(1,0){4}}
      }  
    }
    \multiput(0,0)(8,8){2}{
      \multiput(1,0)(1,0){3}{\line(0,1){4}}
      \multiput(0,1)(0,1){3}{\line(1,0){4}}
    }  
    \multiput(4,4.8)(0,0.8){4}{\line(1,0){4}}
    \multiput(4.8,4)(0.8,0){4}{\line(0,1){4}}
    \multiput(8,8.5)(0,1){4}{\line(1,0){4}}
    \multiput(8.5,8)(1,0){4}{\line(0,1){4}}
  \end{picture}
  \caption{Grid refinement on a cubed-sphere grid mosaic.}
  \label{fig:GridStd:cubeRefined}
\end{figure}
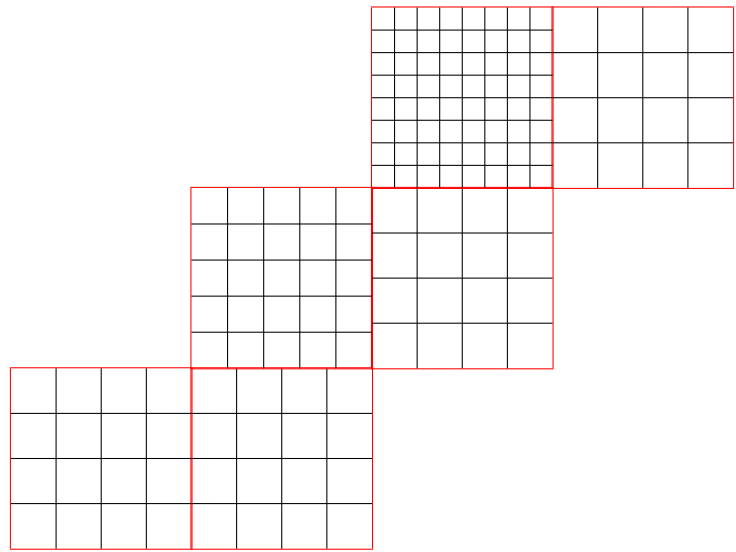

A boundary has the property of \emph{alignment} when there is an
anchor point in \emph{index space} shared by the two grid tiles, i.e
it is possible to state that some point \texttt{(i1,j1)} on grid tile
1 is the same physical point as \texttt{(i2,j2)} on grid tile 2. An
aligned boundary has \emph{no refinement} when the grid lines crossing
the boundary are \emph{continuous}, as in grid tiles 1 and 2 in
\figref{cubeRefined}. The refinement is \emph{integer} when grid lines
from the coarse grid are continuous on the fine grid, but not vice
versa, see grid tiles 5 and 6. The refinement is \emph{rational} in
the example of tile 3, when the contact grid tiles have grid line
counts that are co-prime.

These properties, if present, will aid in the creation of simple and
fast methods for transforming data between grid tiles. If none of the
conditions above are met, there is no alignment. Anchor points are
then represented by geo-referenced coordinates, and remapping is
mediated by an exchange, as described below in \secref{XGrid}.

\subsection{Overlap contact regions: Exchange grids and masks}
\label{sec:GridStd:XGrid}

When there are overlapping grid tiles, the \emph{exchange grid}
construct of \bibref{balajietal2006} is a useful encapsulation of all
the information for conservative interpolation of scalar
quantities.\footnote{Streamline-preserving interpolation of vector
  quantities between grids is still under study, and may result in
  extensions to this proposed grid standard.} The exchange grid,
defined here, does not imply or force any particular algorithm or
conservation requirement; rather it enables conservative regridding of
any order. Methods for creation of exchange grids are briefly
discussed, but the standard is of course divorced from any
implementation.

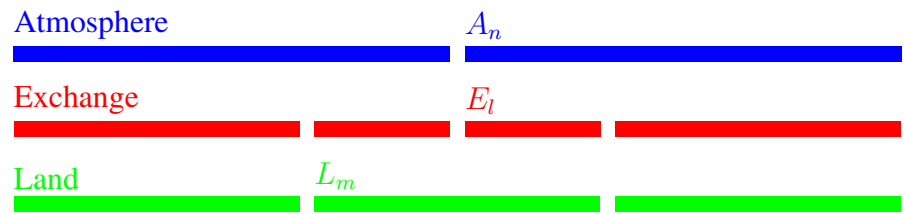
\begin{figure}[htb]
  \setlength{\unitlength}{10mm}
  \begin{center}
    \begin{picture}(12,2.2)
      {\linethickness{2mm}
        \multiput(0,2)(6,0){2}{\color{blue}\line(1,0){5.8}}
        {\color{red}
          \multiput(0,1)(8,0){2}{\line(1,0){3.8}}
          \multiput(4,1)(2,0){2}{\line(1,0){1.8}}}
        \multiput(0,0)(4,0){3}{\color{green}\line(1,0){3.8}}
      }
      \put(0,2.2){\color{blue} \makebox(2,1)[lb]{Atmosphere}}
      \put(0,1.2){\color{red}  \makebox(2,1)[lb]{Exchange}}
      \put(0,0.2){\color{green}\makebox(2,1)[lb]{Land}}
      \put(6,2.2){\color{blue} \makebox(2,1)[lb]{$A_n$}}
      \put(6,1.2){\color{red}  \makebox(2,1)[lb]{$E_l$}}
      \put(4,0.2){\color{green}\makebox(2,1)[lb]{$L_m$}}
    \end{picture}
  \end{center}
  \caption{One-dimensional exchange grid.}
  \label{fig:GridStd:simple}
\end{figure}

Given two grid tiles, an \emph{exchange grid} is the set of
cells defined by the union of all the vertices of the two parent
grid tiles. This is illustrated in~\figref{simple} in 1D, with two parent
grid tiles (``atmosphere'' and ``land''). (\figref{mask} shows an example
of a 2D exchange grid, most often used in practice). As seen here. each
exchange grid cell can be uniquely associated with exactly one cell on
each parent grid tile, and \emph{fractional areas} with respect to the
parent grid cells. Quantities being transferred from one parent grid tile to
the other are first interpolated onto the exchange grid using one set
of fractional areas; and then averaged onto the receiving grid using
the other set of fractional areas. If a particular moment of the
exchanged quantity is required to be conserved, consistent
moment-conserving interpolation and averaging functions of the
fractional area may be employed. This may require not only the
cell-average of the quantity (zeroth-order moment) but also
higher-order moments to be transferred across the exchange grid.

Given $N$ cells of one parent grid tile, and $M$ cells of the other,
the exchange grid is, in the limiting case in which every cell on one
grid overlaps with every cell on the other, a matrix of size $N\times
M$. In practice, however, very few cells overlap, and the exchange
grid matrix is extremely sparse. In code, we typically treat the
exchange grid cell array as a compact 1D array (thus shown
in~\figref{simple} as \(E_l\) rather than \(E_{nm}\)) with indices
pointing back to the parent grid tile cells. \tabref{xgrid} shows the
characteristics of exchange grids at typical climate model
resolutions. The first is the current GFDL model CM2
\citep{ref:delworthetal2006}, and the second for a projected
next-generation model still under development. As seen here, the
exchange grids are extremely sparse.

\begin{table}[htb]
  \begin{center}
    \begin{tabular}[c]{{||c|c|c|c|c||}} \hline
      Atmosphere & Ocean & Xgrid & Density & Scalability \\ \hline\hline
      144$\times$90 & 360$\times$200 & 79644 & $8.5 \times 10^{-5}$ & 0.29
      \\ \hline
      288$\times$180 & 1080$\times$840 & 895390 & $1.9 \times 10^{-5}$ &
      0.56 \\ \hline
    \end{tabular}
  \end{center}
  \caption{Exchange grid sizes for typical climate model grids. The
    first column shows the horizontal discretization of an atmospheric
    model at ``typical" climate resolutions of 2\degree and 1\degree
    respectively. The ``ocean" column shows the same for an ocean
    model, at 1\degree and $\frac13\degree$. The ``Xgrid" column shows
    the number of points in the computed exchange grid, and the
    density relates that to the theoretical maximum number of exchange
    grid cells. The ``scalability" column shows the load imbalance of
    the exchange grid relative to the overall model when it inherits
    its parallel decomposition from one of the parent grid tiles.}
  \label{tab:GridStd:xgrid}
\end{table}
  
The computation of the exchange grid itself could be time consuming,
for parent grid tiles on completely non-conformant curvilinear
coordinates. In practice, this issue is often sidestepped by
precomputing and storing the exchange grid. The issue must be
revisited if either of the parent grid tiles is adaptive. Methods for
exchange grid computation include the
\href{http://climate.lanl.gov/Software/SCRIP}{SCRIP} package
\citep{ref:jones1999} and others based on discretizing the underlying
continuous geometry as a raster of high-resolution pixels
\citep{ref:kosteretal2000}.

This illustration of exchange grids restricts itself to 2-dimensional
LRGs on the planetary surface. However, there is nothing in the
exchange grid concept that prevents its use in any of the
discretizations of \secref{Discretization}, or in exchanges between
grids varying in 3, or even 4 (including time) dimensions.

\subsubsection{Masks}

A complication arises when one of the surfaces is partitioned into
\emph{complementary components}: in Earth system models, a typical
example is that of an ocean and land surface that together tile the
area under the atmosphere. Conservative exchange between \emph{three}
components may then be required: quantities like CO\(_2\) have
reservoirs in all three media, with the \emph{total} carbon inventory
being conserved.

\begin{figure}[htb]
  \setlength{\unitlength}{10mm}
  \begin{center}
    \begin{picture}(12,5)
      \thicklines
      \put(0,0){\framebox(3,1){Land}}
      \put(4,0){\framebox(3,1){Ocean}}
      \put(8,0){\framebox(3,1){Exchange}}

      \multiput(0,2)(1,0){4}{\line(0,1){3}}
      \multiput(0,2)(0,1){4}{\line(1,0){3}}
      \put(0,3){\makebox(1,1){\textbf{L}}}
      \put(0,4){\makebox(1,1){\textbf{L}}}
      \put(1,4){\makebox(1,1){\textbf{L}}}

      \multiput(4,2)(1.5,0){3}{\line(0,1){3}}
      \multiput(4,2)(0,1.5){3}{\line(1,0){3}}
      \put(4,3.5){\makebox(1.5,1.5){\textbf{L}}}
      
      \multiput(8,2)(1,0){4}{\line(0,1){3}}
      \multiput(8,2)(0,1){4}{\line(1,0){3}}
      \put(8,3.5){\line(1,0){3}}
      \put(9.5,2){\line(0,1){3}}
      
      \linethickness{0.5mm} 
      \put(-0.1,0){
        \color{green}
        \put(4,3.5){\line(1,0){1.5}}
        \put(5.5,3.5){\line(0,1){1.5}}

        \multiput(0,3)(1,1){2}{\line(1,0){1}}
        \multiput(1,3)(1,1){2}{\line(0,1){1}}

        \color{blue}
        \put(8,3){\framebox(1,0.5){}}
        \put(9.5,4){\framebox(0.5,1){}}
        \color{red}
        \put(9,3.5){\framebox(0.5,0.5){}}
      }
    \end{picture}
  \end{center}  
  \caption{The mask problem. The land and atmosphere share the grid on
    the left, and their discretization of the land-sea mask is
    different from the ocean model, in the middle. The exchange
    grid, right, is where these may be reconciled: the red
    ``orphan" cell is assigned (arbitrarily) to the land, and the
    land cell areas ``clipped" to remove the doubly-owned blue cells.}
  \label{fig:GridStd:mask}
\end{figure}
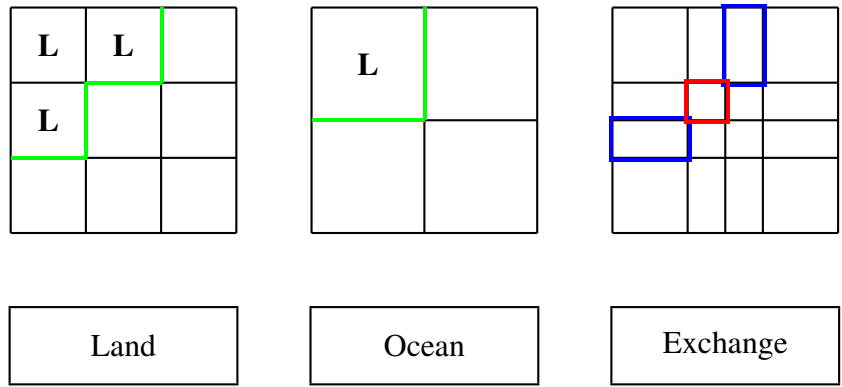

\figref{mask} shows such an instance, with an atmosphere-land grid and
an ocean grid of different resolution. The green line in the first two
frames shows the \emph{land-sea mask} as discretized on the two grids,
with the cells marked \textbf{L} belonging to the land. Due to the
differing resolution, certain exchange grid cells have ambiguous
status: the two blue cells are claimed by both land and ocean, while
the orphan red cell is claimed by neither.

This implies that the mask defining the boundary between complementary
grids can only be accurately defined on the exchange grid: only there
can it be guaranteed that the cell areas exactly tile the global
domain. Cells of ambiguous status are resolved here, by adopting some
ownership convention. For example, in the FMS exchange grid, we
generally modify the land model as needed: the land grid cells
are quite independent of each other and amenable to such
transformations. We add cells to the land grid until there are no
orphan ``red'' cells left on the exchange grid, then get rid of the
``blue'' cells by \emph{clipping} the fractional areas on the land
side.

\section{Representing the grid vocabulary in the CF conventions}
\label{sec:GridStd:CF}

The CF conventions have been developed in the context of the netCDF
data format. The current momentum is toward using technologies such as
OpenDAP to achieve format neutrality for data; and to develop the
conventions themselves toward a standard through a mechanism such as
OGC. As the standardization process continues, it is likely that much
of CF metadata will be stored in databases in a readily-harvested form
such as XML. For the purposes of this paper, however, we will continue
to represent the contents of the grid standard using netCDF
terminology, as now.

The current CF standard covers data fields for single grid tiles very
well. As there are considerable data archives already storing data in
this form, we have tried to do the least violence to existing data
representations of variables on single grid tiles. The proposed
extensions serve as enhancements to CF that will allow a full
expression for data discretized on grid mosaics. Features to highlight
include:

\begin{itemize}
\item a standard grid specification dataset (or \emph{gridspec}) for
  grid mosaics. The grid specification is comprehensive and is
  potentially a very large file. Various CF attributes will be used to
  indicate properties of the grid that permit a succinct description
  from which the complete gridspec is readily reconstructed.
\item an extended family of CF standard names for grid specification;
\item netCDF and CF currently assume that all information is present
  in a single \emph{file}. This assumption is already currently broken
  in many ways: for instance it is customary to store a long time
  series of a variable in multiple files. The assumption is also often
  flawed for vector fields: vector components may be stored as
  multiple files. We propose here a mechanism for storing a
  CF-compliant \emph{dataset} in multiple \emph{files}\footnote{The
    HDF5 specification, with which netCDF will merge, takes a
    filesystem-within-a-file approach to this problem, which by all
    accounts is not very efficient \missref{}. The proposed approach
    will allow very efficient approaches to dataset aggregation.}, and
  for preserving (or at least verifying) integrity of a multi-file
  dataset.
\item The gridspec is a work in progress, and is designed for
  extensibility. We expect to see considerable evolution in the near
  term. It is therefore liberally sprinkled with \emph{version}
  metadata.
\end{itemize}

The general approach is as follows. Datasets are generally archived in
a way whereby one approaches the dataset following metadata that
describes the experiment to which it belongs. The gridspec forms part
of the experiment metadata. For Earth System models, comprehensive
model metadata is under development. A gridspec describing the
complete grid mosaic of an entire coupled model (shown schematically
in \figref{mosaicTree}) will be stored under the experiment, and we
expect software processing any dataset associated with the experiment
to have access to the gridspec.\footnote{As the gridspec is also
  intended for use as model input, said software might indeed be an
  Earth system model.}

Datasets holding physical variables will not themselves refer to the
gridspec; the connection is made at the metadata level above.

Physical variables discretized on a mosaic of more than one grid tile may
be stored in multiple files, where each file contains one or more grid
tiles.

\subsection{Linkages between files}
\label{sec:GridStd:LinkSpec}

We propose that links be directed and acyclic: e.g grid mosaic files
point to constituent grid tile files, but the ``leaf'' files do not
point back.

Files may be described using local pathnames or remote URIs (URLs,
OpenDAP IDs). File descriptors may be absolute or relative to a base
address, as in HTML.

When pointing to an external file, attributes holding the timestamp
and MD5 checksum\footnote{MD5 checksums are standard practice. One can
  intentionally generate, by bit exchanges, erroneous files that give
  the same MD5 checksum, but the probability of this occurring by
  coincidence is vanishingly small. MD5 checksums have been measured
  to take about a minute for a 10Gb dataset.} may optionally be
specified. If the checksum of an external file does not match, it is
an error. The timestamp is not definitive, but may be used to decide
whether or not to trigger a checksum.

\codeBlock{GridStd:LinkSpec}{
  dimensions: \\
  \qquad string = 255; \\
  variables: \\
  \qquad char base(string); \\
  \qquad char external(string); \\
  \qquad char local(string); \\
  base = "http://www.gfdl.noaa.gov/CM2.1/"; \\
  \qquad base:standard_name = "link_base_path"; \\
  external = "foo.nc"; \\
  \qquad external:standard_name = "link_path"; \\
  \qquad external:md5_checksum = "g0bbl3dyg00k"; \\
  \qquad external:timestamp = "20060509T012800.33Z"; \\
  local = "/home/foo/bar.nc"; \\
  \qquad local:standard_name = "link_path"; \\
  \qquad local:link_spec_version = "0.2";
}

Encoding pathnames, checksums and timestamps carries a penalty: the
system is brittle to any changes. The use of relative pathnames is
recommended: this at least permits whole directory trees to be moved
with little pain.

\noindent \textbf{Summary}: two new standard names
\texttt{link_base_path} and \texttt{link_path}. Optional attributes:
\texttt{link_spec_version}, \texttt{md5_checksum} and
\texttt{timestamp}.

\subsection{Grid mosaic}
\label{sec:GridStd:MosaicSpec}

The grid mosaic specification is identified by a unique string name
which qualifies its interior namespace. As shown schematically in
\figref{mosaicTree}, its children can be mosaics or grid tiles.
Contact regions are specified between pairs of grid tiles, using the
fully qualified grid tile specification \emph{mosaic:mosaic:...:tile}.

\codeBlock{GridStd:MosaicSpec}{
  dimensions: \\
  \qquad nfaces = 6; \\
  \qquad ncontact = 12; \\
  \qquad string = 255; \\
  variables: \\
  \qquad char mosaic(string); \\
  \qquad char gridfaces(nfaces,string); \\
  \qquad char contacts(ncontact,string); \\
  mosaic = "AM2C45L24"; \\
  \qquad mosaic:standard_name = "grid_mosaic_spec"; \\
  \qquad mosaic:mosaic_spec_version = "0.2"; \\
  \qquad mosaic:children = "gridfaces"; \\
  \qquad mosaic:contact_regions = "contacts"; \\
  \qquad mosaic:grid_descriptor = "C45L24 cubed_sphere"; \\
  gridfaces = \\
  \qquad "Face1", \\
  \qquad "Face2", \\
  \qquad "Face3", \\
  \qquad "Face4", \\
  \qquad "Face5", \\
  \qquad "Face6"; \\
  contacts = \\
  \qquad "AM2C45L24:Face1::AM2C45L24:Face2", \\
  \qquad "AM2C45L24:Face1::AM2C45L24:Face3", \\
  \qquad "AM2C45L24:Face1::AM2C45L24:Face5", \\
  \qquad "AM2C45L24:Face1::AM2C45L24:Face6", \\
  \qquad "AM2C45L24:Face2::AM2C45L24:Face3", \\
  \qquad "AM2C45L24:Face2::AM2C45L24:Face4", \\
  \qquad "AM2C45L24:Face2::AM2C45L24:Face6", \\
  \qquad "AM2C45L24:Face3::AM2C45L24:Face4", \\
  \qquad "AM2C45L24:Face3::AM2C45L24:Face5", \\
  \qquad "AM2C45L24:Face4::AM2C45L24:Face5", \\
  \qquad "AM2C45L24:Face4::AM2C45L24:Face6", \\
  \qquad "AM2C45L24:Face5::AM2C45L24:Face6";
}

\noindent \textbf{Summary}: a new standard names
\texttt{grid_mosaic_spec}. Grid mosaic specs have
attributes \texttt{mosaic_spec_version}, \texttt{children}
and \texttt{contact_regions}. Optional attributes
\texttt{children_links} and \texttt{contact_region_links} may point to
external files containing the specifications for the children and
their contacts.

The \emph{grid_descriptor} is an optional text description of the grid
that uses commonly used terminology, but may not in general be a sufficient
description of the field (many grids are numerically generated, and do
not admit of a succinct description). Examples of grid descriptors
include:

\begin{itemize}
\item \texttt{spectral_gaussian_grid}
\item \texttt{regular_lon_lat_grid}
\item \texttt{reduced_gaussian_grid}
\item \texttt{displaced_pole_grid} (different from a \emph{rotated
    pole grid}: any grid could have a rotated north pole);
\item \texttt{tripolar_grid}
\item \texttt{cubed_sphere_grid}
\item \texttt{icosahedral_geodesic_grid}
\item \texttt{yin_yang_grid}
\end{itemize}

The grid descriptor could additionally contain common shorthand
descriptions such as \texttt{t42}, or perhaps could go further toward
machine processing using terms like \texttt{triangular_truncation}.

\subsection{Grid tile}
\label{sec:GridStd:TileSpec}

\codeBlock{GridStd:GridTileSpec}{
  dimensions: \\
  \qquad string = 255; \\
  \qquad nx = 90; \\
  \qquad ny = 90; \\
  \qquad nxv = 91; \\
  \qquad nyv = 91; \\
  \qquad nz = 24; \\
  variables: \\
  \qquad char tile(string); \\
  \qquad\qquad tile:standard_name = "grid_tile_spec"; \\
  \qquad\qquad tile:tile_spec_version = "0.2"; \\
  \qquad\qquad tile:geometry = "spherical"; \\
  \qquad\qquad tile:north_pole = "0.0 90.0"; \\
  \qquad\qquad tile:projection = "cube_gnomonic"; \\
  \qquad\qquad tile:discretization = "logically_rectangular"; \\
  \qquad\qquad tile:conformal = "true"; \\
  \qquad double area(ny,nx); \\
  \qquad\qquad area:standard_name = "grid_cell_area"; \\
  \qquad\qquad area:units = "m\^2"; \\
  \qquad double dx(ny+1,nx); \\
  \qquad\qquad dx:standard_name = "grid_edge_x_distance"; \\
  \qquad\qquad dx:units = "metres"; \\
  \qquad double dy(ny,nx+1); \\
  \qquad\qquad dy:standard_name = "grid_edge_y_distance"; \\
  \qquad\qquad dy:units = "metres"; \\
  \qquad double angle_dx(ny+1,nx); \\
  \qquad\qquad angle_dx:standard_name = \\
  \qquad\qquad\qquad "grid_edge_x_angle_WRT_geographic_east"; \\
  \qquad\qquad angle_dx:units = "radians"; \\
  \qquad char arcx(string); \\
  \qquad\qquad arcx:standard_name = "grid_edge_x_arc_type"; \\
  \qquad double zeta(nz); \\
  arcx = "great_circle"; \\
  tile = "Face1";
}

Horizontal vertex location specifications may be of different rank
depending on their regularity or uniformity. (Note that the
geo-referencing information may still be 2D even for regular coordinates).

An irregular horizontal grid requires a 2D specification of vertex
locations:

\codeBlock{GridStd:Regular}{
  variables: \\
  \qquad float geolon(ny+1,nx+1); \\
  \qquad\qquad geolon:standard_name = "geographic_longitude"; \\
  \qquad float geolat(ny+1,nx+1); \\
  \qquad\qquad geolat:standard_name = "geographic_latitude"; \\
  \qquad float xvert(ny+1,nx+1); \\
  \qquad\qquad xvert:standard_name = "grid_longitude"; \\
  \qquad\qquad xvert:geospatial_coordinates = "geolon geolat"; \\
  \qquad float yvert(ny+1,nx+1); \\
  \qquad\qquad yvert:standard_name = "grid_latitude"; \\
  \qquad\qquad yvert:geospatial_coordinates = "geolon geolat";
}

The vertical geo-mapping is expressed by reference to ``standard
levels''. 

\noindent \textbf{Summary}: several new standard names to describe
properties of a grid: distances, angles, areas and volumes. The
\emph{arc type} is a new variable with no equivalent in CF. Currently,
we are considering values of \texttt{great_circle} and
\texttt{small_circle}, but others may be imagined. The
\texttt{small_circle} arc type requires the specification of a pole.

The grid tile spec has attributes geometry (\secref{Geometry}),
projection (\secref{HorizCoord}: a value of \texttt{none} indicates no
projection) and discretization (\secref{Discretization}). The optional
attributes \texttt{regular}, \texttt{conformal} and \texttt{uniform}
may be used to shrink the grid tile spec.

\subsection{Unstructured grid tile}
\label{sec:GridStd:Unstructured}

The unstructured grid tile is an UTG. The current specification
follows an actual example used by the FVCOM model \missref{Gross;
  Signell}. While in the LRG example above, the number of vertices can
be deduced from the number of cells, it cannot in the unstructured
case.

Each cell is modeled as triangular. Distances, arc types, angles and
areas are cell properties. Additional elements of the UTG
specification are variables with standard names of
\texttt{vertex_index} and \texttt{neighbor_cell_index} to contain the
indices of a cell's 3 vertices and its 3 neighbours, respectively. The
ordering line segments, neighbors, etc., all follow the ordering of
vertices.

\codeBlock{GridStd:UTGSpec}{
  dimensions: \\
  \qquad string = 255; \\
  \qquad node = 871; \\
  \qquad nele = 1620; \\
  variables: \\
  \qquad char tile(string); \\
  \qquad\qquad tile:standard_name = "grid_tile_spec"; \\
  \qquad\qquad tile:tile_spec_version = "0.2"; \\
  \qquad\qquad tile:geometry =  "spherical"; \\
  \qquad\qquad tile:north_pole = "0.0 90.0"; \\
  \qquad\qquad tile:discretization = \\
  \qquad\qquad\qquad "unstructured_triangular"; \\
  \qquad double area(nele); \\
  \qquad\qquad area:standard_name = "grid_cell_area"; \\
  \qquad\qquad area:units = "m\^2"; \\
  \qquad double ds(3,nele); \\
  \qquad\qquad ds:standard_name = "grid_edge_distance"; \\
  \qquad\qquad ds:units = "metres"; \\
  \qquad double angle_ds(3,nele); \\
  \qquad\qquad angle_ds:standard_name = \\
  \qquad\qquad\qquad "grid_edge_angle_WRT_geographic_east"; \\
  \qquad\qquad angle_ds:units = "radians"; \\
  \qquad char arcx(string); \\
  \qquad\qquad arcx:standard_name = "grid_edge_arc_type"; \\
  \qquad int nv(3,nele); \\
  \qquad\qquad nv:standard_name = "neighbor_cell_index"; \\
  \qquad int node_index(3,nele); \\
  \qquad\qquad node_index:standard_name = "vertex_index"; \\
  arcx = "great_circle"; \\
  tile = "fvcom_grid";
}

\codeBlock{GridStd:UTGCoords}{
  variables: \\
  \qquad float geolon(node); \\
  \qquad\qquad geolon:standard_name = "geographic_longitude"; \\
  \qquad float geolat(node); \\
  \qquad\qquad geolat:standard_name = "geographic_latitude"; \\
  \qquad float xvert(node); \\
  \qquad\qquad xvert:standard_name = "grid_x_coordinate"; \\
  \qquad\qquad xvert:units = "metres"; \\
  \qquad\qquad xvert:geospatial_coordinates = "geolon geolat"; \\
  \qquad float yvert(node); \\
  \qquad\qquad yvert:standard_name = "grid_y_coordinate"; \\
  \qquad\qquad yvert:units = "metres"; \\
  \qquad\qquad yvert:geospatial_coordinates = "geolon geolat";
}


\subsection{Contact regions}
\label{sec:GridStd:ContactSpec}

\codeBlock{GridStd:GridBoundarySpec}{
  dimensions: \\
  \qquad string = 255; \\
  variables: \\
  \qquad int anchor(2,2); \\
  \qquad\qquad anchor:standard_name = \\
  \qquad\qquad\qquad "anchor_point_shared_between_tiles"; \\
  \qquad char orient(string); \\
  \qquad\qquad orient:standard_name = \\
  \qquad\qquad\qquad "orientation_of_shared_boundary"; \\
  \qquad char contact(string); \\
  \qquad\qquad contact:standard_name = "grid_contact_spec"; \\
  \qquad\qquad contact:contact_spec_version = "0.2"; \\
  \qquad\qquad contact:contact_type = "boundary"; \\
  \qquad\qquad contact:alignment = "true"; \\
  \qquad\qquad contact:refinement = "none"; \\
  \qquad\qquad contact:anchor_point = "anchor"; \\
  \qquad\qquad contact:orientation = "orient"; \\
  contact = "AM2C45L24:Face1::AM2C45L24:Face2"; \\
  orient = "Y:Y"; \\
  anchor = "90 1 1 1";
}

\codeBlock{GridStd:GridExchangeSpec}{
  dimensions: \\
  \qquad string = 255; \\
  \qquad ncells = 1476; \\
  variables: \\
  \qquad double frac_area(2,ncells); \\
  \qquad\qquad frac_area:standard_name = \\
  \qquad\qquad\qquad "fractional_area_of_exchange_grid_cell"; \\
  \qquad int tile1_cell(2,ncells); \\
  \qquad\qquad tile1_cell:standard_name="parent_cell_indices"; \\
  \qquad int tile2_cell(2,ncells); \\
  \qquad\qquad tile2_cell:standard_name="parent_cell_indices"; \\
  \qquad char contact(string); \\
  \qquad\qquad contact:standard_name = "grid_contact_spec"; \\
  \qquad\qquad contact:contact_spec_version = "0.2"; \\
  \qquad\qquad contact:contact_type = "exchange"; \\
  \qquad\qquad contact:fractional_area_field = "frac_area"; \\
  \qquad\qquad contact:parent1_cell = "tile1_cell"; \\
  \qquad\qquad contact:parent2_cell = "tile2_cell"; \\
  contact = "CM2:LM2::AM2C45L24:Face2";
}

\subsection{Variables}
\label{sec:GridStd:Field}

Variables are held in CF-compliant files that are separate from the
gridspec but can link to it following the link spec in
\secref{LinkSpec}. Variables on a single grid tile can follow CF-1.0,
with no changes. The additional information provided by the gridspec
can be linked in, as shown in this example of a $U$ velocity component
on a C grid (\figref{cgrid}).

\codeBlock{GridStd:Field}{
  dimensions: \\
  \qquad nx = 46; \\
  \qquad ny = 45; \\
  variables: \\
  \qquad int nx_u(nx); \\
  \qquad int ny_u(ny); \\
  \qquad float u(ny,nx); \\
  \qquad\qquad u:standard_name = "grid_eastward_velocity"; \\
  \qquad\qquad u:staggering = "c_grid_symmetric"; \\
  GLOBAL ATTRIBUTES: \\
  \qquad gridspec = "foo.nc"; \\
  nx_u = 1,3,5,... \\
  ny_u = 2,4,6,...
}

The \emph{staggering} field expresses what is implicit in the values
of \texttt{nx_u} and \texttt{ny_u}, but is useful
nonetheless\footnote{In general, there may be a lot of redundancy in
  the gridspec, which poses a \emph{consistency} problem. In general,
  consistency checking and validation are relatively simple, as in the
  instance here.}. Possible values of \texttt{staggering} include:

\begin{itemize}
\item \texttt{c_grid_symmetric}
\item \texttt{c_grid_ne}
\item \texttt{b_grid_sw}
\item ... and so on.
\end{itemize}

Using this information, it is possible to perform correct
transformations, such as combining this field with a $V$ velocity from
another file, transforming to an A-grid, and then rotating to
geographic coordinates.

\section{Examples}
\label{sec:GridStd:Examples}

\subsection{Cartesian geometry}
\label{sec:GridStd:cartesian}

\codeBlock{GridStd:Cartesian}{
  dimensions: \\
  \qquad string = 255; \\
  \qquad nx = 8; \\
  \qquad ny = 8; \\
  variables: \\
  \qquad char tile(string); \\
  \qquad\qquad tile:standard_name = "grid_tile_spec"; \\
  \qquad\qquad tile:tile_spec_version = "0.2"; \\
  \qquad\qquad tile:geometry = "planar"; \\
  \qquad\qquad tile:projection = "cartesian"; \\
  \qquad\qquad tile:discretization = "logically_rectangular"; \\
  \qquad\qquad tile:conformal = "true"; \\
  \qquad\qquad tile:uniform = "true"; \\
  \qquad double area; \\
  \qquad\qquad area:standard_name = "grid_cell_area"; \\
  \qquad\qquad area:units = "m\^2"; \\
  \qquad double dx; \\
  \qquad\qquad dx:standard_name = "grid_edge_x_distance"; \\
  \qquad\qquad dx:units = "metres"; \\
  \qquad double dy; \\
  \qquad\qquad dy:standard_name = "grid_edge_y_distance"; \\
  \qquad\qquad dy:units = "metres"; \\
  tile = "Descartes";
}


The Cartesian grid spec of \ceqref{Cartesian} illustrates several
simplifications with respect to \ceqref{GridTileSpec}.

\begin{itemize}
\item The \texttt{geometry:planar} attribute (\secref{Geometry})
  indicates that geo-referencing is not possible.
\item Since the \texttt{uniform} attribute (\secref{HorizCoord}) is
  set, the \texttt{area}, \texttt{dx} and \texttt{dy} fields reduce to
  simple scalars.
\item The combination of a conformal attribute and the planar geometry
  means that it is not required to store angles: grid lines are
  orthogonal, and that's that.
\item The tile name is of course arbitrary: we have chosen to type the
  tile as a string to avoid using the derived or complex types of
  \href{http://www.unidata.ucar.edu/software/netcdf/netcdf-4}{netCDF-4}.
  Mosaic processing tools will enforce the absence of two tiles
  bearing the same name.
\end{itemize}

Note that this gridspec might actually represent a supergrid of a
4$\times$4 grid: we cannot tell from the gridspec alone. We would need
to examine a field containing a physical variable (\secref{Field}).

\subsection{Gaussian grid}
\label{sec:GridStd:gaussian}

\codeBlock{GridStd:Gaussian}{
  dimensions: \\
  \qquad string = 255; \\
  \qquad nx = 320; \\
  \qquad ny = 160; \\
  variables: \\
  \qquad char tile(string); \\
  \qquad\qquad tile:standard_name = "grid_tile_spec"; \\
  \qquad\qquad tile:tile_spec_version = "0.2"; \\
  \qquad\qquad tile:geometry = "spherical"; \\
  \qquad\qquad tile:north_pole = "0.0 90.0"; \\
  \qquad\qquad tile:discretization = "logically_rectangular"; \\
  \qquad\qquad tile:horizontal_grid_descriptor = "gaussian_grid"; \\
  \qquad\qquad tile:conformal = "true"; \\
  \qquad\qquad tile:regular = "true"; \\
  \qquad double area(ny,nx); \\
  \qquad\qquad area:standard_name = "grid_cell_area"; \\
  \qquad\qquad area:units = "m\^2"; \\
  \qquad double dx(nx); \\
  \qquad\qquad dx:standard_name = "grid_edge_x_distance"; \\
  \qquad\qquad dx:units = "metres"; \\
  \qquad double dy(ny); \\
  \qquad\qquad dy:standard_name = "grid_edge_y_distance"; \\
  \qquad\qquad dy:units = "metres"; \\
  \qquad double angle_dx(,nx); \\
  \qquad\qquad angle_dx:standard_name = \\
  \qquad\qquad\qquad "grid_edge_x_angle_WRT_geographic_east"; \\
  \qquad\qquad angle_dx:units = "radians"; \\
  \qquad char arcx(string); \\
  \qquad\qquad arcx:standard_name = "grid_edge_x_arc_type"; \\
  \qquad double zeta(nz); \\
  arcx = "small_circle"; \\
  tile = "T106";
}

\codeBlock{GridStd:GaussianContact}{
  dimensions: \\
  \qquad string = 255; \\
  variables: \\
  \qquad int anchor(2,2); \\
  \qquad\qquad anchor:standard_name = \\
  \qquad\qquad\qquad "anchor_point_shared_between_tiles"; \\
  \qquad char orient(string); \\
  \qquad\qquad orient:standard_name = \\
  \qquad\qquad\qquad "orientation_of_shared_boundary"; \\
  \qquad char contact(string); \\
  \qquad\qquad contact:standard_name = "grid_contact_spec"; \\
  \qquad\qquad contact:contact_spec_version = "0.2"; \\
  \qquad\qquad contact:contact_type = "boundary"; \\
  \qquad\qquad contact:alignment = "true"; \\
  \qquad\qquad contact:refinement = "none"; \\
  \qquad\qquad contact:anchor_point = "anchor"; \\
  \qquad\qquad contact:orientation = "orient"; \\
  contact = "Gaussian::Gaussian"; \\
  orient = "Y:Y"; \\
  anchor = "320 1 1 1";
}

A Gaussian grid is a spatial grid where locations on a sphere are
generated by ``Gaussian quadrature'' from a given truncation of
spherical harmonics in spectral space.

\begin{itemize}
\item There is no projection onto a plane.
\item Since this is a \texttt{regular} grid (\secref{HorizCoord}),
  \texttt{dx} and \texttt{dy} are 1D rather than 2D arrays. The
  specification of angles is similarly reduced by the
  \texttt{conformal} attribute.
\item The contact spec in \ceqref{GaussianContact} specifies
  periodicity in $X$.
\item The associated mosaic specification is not shown here, as a
  simple Gaussian grid is a mosaic of a single tile. The
  \texttt{horizontal_grid_descriptor} (\secref{MosaicSpec}) is given a
  value of \texttt{spectral_gaussian_grid}: this value belongs to a
  \texttt{controlled vocabulary} of grid descriptors. The combination
  of this descriptor with the truncation level is enough to completely
  specify the gaussian grid.
\end{itemize}

\subsection{Reduced gaussian grid}
\label{sec:GridStd:reduced}

A Gaussian grid is of course a kind of \texttt{regular_lat_lon_grid},
and can suffer from various numerical problems owing to the
convergence of longitudes near the poles. The \emph{reduced} Gaussian
grid of \bibref{hortalsimmons1991} overcomes this problem by reducing
the number of longitudes within latitute bands approaching the pole,
as shown in \figref{reduced}.

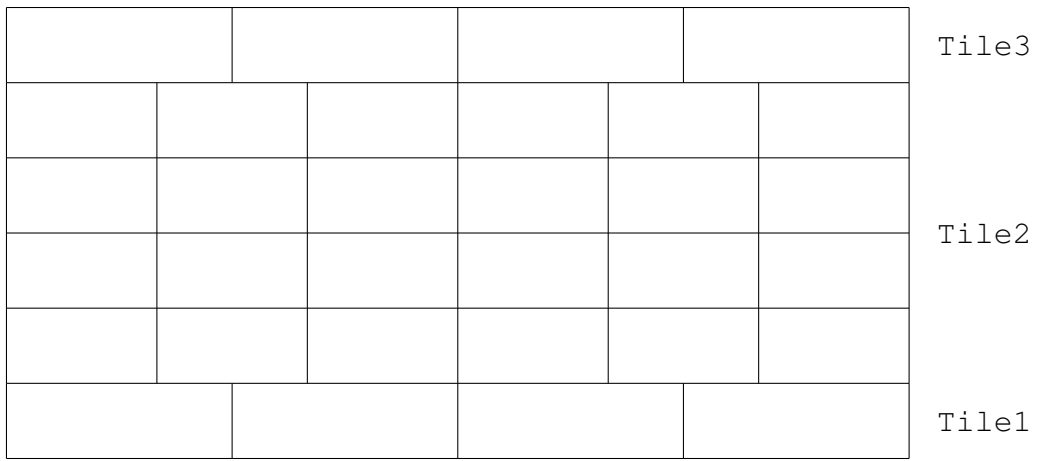
\begin{figure}[htb]
  \setlength{\unitlength}{10mm}
  \begin{center}
    \begin{picture}(14,6)
      \multiput(0,0)(0,1){7}{\line(1,0){12}}
      \multiput(0,0)(0,5){2}{
        \multiput(0,0)(3,0){5}{\line(0,1){1}}
      }
      \multiput(0,1)(2,0){7}{\line(0,1){4}}
      \put(12,0){\makebox(2,1){\texttt{Tile1}}}
      \put(12,1){\makebox(2,4){\texttt{Tile2}}}
      \put(12,5){\makebox(2,1){\texttt{Tile3}}}
    \end{picture}
  \end{center}
  \caption{Reduced Gaussian grid.}
  \label{fig:GridStd:reduced}
\end{figure}

\codeBlock{GridStd:ReducedGaussian}{
  dimensions: \\
  \qquad ntiles = 6; \\
  \qquad ncontact = 5; \\
  \qquad string = 255; \\
  variables: \\
  \qquad char mosaic(string); \\
  \qquad char gridtiles(nfaces,string); \\
  \qquad char contacts(ncontact,string); \\
  mosaic = "Hortal"; \\
  \qquad mosaic:standard_name = "grid_mosaic_spec"; \\
  \qquad mosaic:mosaic_spec_version = "0.2"; \\
  \qquad mosaic:children = "gridfaces"; \\
  \qquad mosaic:contact_regions = "contacts"; \\
  \qquad mosaic:grid_descriptor = "reduced_gaussian_grid"; \\
  gridtiles = \\
  \qquad "Tile1", \\
  \qquad "Tile2", \\
  \qquad "Tile3"; \\
  contacts = \\
  \qquad "Hortal:Tile1::Hortal:Tile1", \\
  \qquad "Hortal:Tile2::Hortal:Tile2", \\
  \qquad "Hortal:Tile3::Hortal:Tile3", \\
  \qquad "Hortal:Tile1::Hortal:Tile2", \\
  \qquad "Hortal:Tile2::Hortal:Tile3"; \\
  \dots \\
  contact = "Hortal:Tile1::Hortal:Tile1"; \\
  orient = "Y:Y"; \\
  anchor = "1 1 5 1"; \\
  contact = "Hortal:Tile2::Hortal:Tile2"; \\
  orient = "Y:Y"; \\
  anchor = "1 1 7 1"; \\
  contact = "Hortal:Tile3::Hortal:Tile3"; \\
  orient = "Y:Y"; \\
  anchor = "1 1 5 1"; \\
  contact = "Hortal:Tile1::Hortal:Tile2"; \\
  orient = "X:X"; \\
  anchor = "1 2 1 1"; \\
  contact = "Hortal:Tile2::Hortal:Tile3"; \\
  orient = "X:X"; \\
  anchor = "1 5 1 1";
}

The reduced Gaussian grid of \figref{reduced} is represented as a
mosaic of multiple grid tiles, each of which is restricted to a
latitude band, and has different longitudinal resolution.

\begin{itemize}
\item The mosaic as a whole has the \texttt{reduced_gaussian_grid}
  descriptor.
\item It consists of 3 tiles, as shown in \figref{reduced}, and 5
  contact regions. The first 3 contacts express periodicity in $X$
  within a tile; the last two express contacts between tiles at the
  latitude where the zonal resolution changes. 
\end{itemize}

\subsection{Tripolar grid}
\label{sec:GridStd:tripolar}

The tripolar grid of \figref{tripolar} is a LRG mosaic consisting of a
single tile. The tile is in contact with itself in the manner of a
sheet of paper folded in half. In the $X$ direction, we have simple
periodicity. Along the north edge, there is a fold, which is best
conceived of a boundary in contact with itself with reversed
orientation. Thus, given a tripolar grid called \texttt{murray} of
$M\times N$ points, we would have:

\codeBlock{Gridstd:tripolar}{
  contact = "murray::murray X"; \\
  orient = "Y:Y"; \\
  anchor = "1 M 1 1"; \\
  contact = "murray::murray Y"; \\
  orient = "X:-X"; \\
  anchor = "1 N M N;
}

\subsection{Unstructured triangular grid}
\label{sec:GridStd:unstructured}

We show here an example of fields on a UTG following the FVCOM
example of \secref{Unstructured}. The example shows vertex-centred
scalars and cell-centered velocities:

\codeBlock{GridStd:UTGField}{
  variables: \\
  \qquad float u(nele); \\
  \qquad\qquad u:standard_name = "eastward_velocity"; \\
  \qquad\qquad u:staggering = "cell_centred"; \\
  \qquad float v(nele); \\
  \qquad\qquad v:standard_name = "northward_velocity"; \\
  \qquad\qquad v:staggering = "cell_centred"; \\
  \qquad float t(node); \\
  \qquad\qquad t:standard_name = "temperature"; \\
  \qquad\qquad t:staggering = "vertex_centred"; \\
  GLOBAL ATTRIBUTES: \\
  \qquad gridspec = "foo.nc";
}

\section{Gridspec implementations}
\label{sec:GridStd:Implementations}

There are two pioneering implementations of the Mosaic Gridspec. One
is a complete XML schema developed on the basis of the Gridspec; the
other is a complete netCDF-3 implementation.

\subsection{The GENIE Gridspec}
\label{sec:GridStd:Genie}

The \href{http://www.genie.ac.uk}{GENIE project} has the objective of
building a Grid-based Earth system model that will built out of
component models drawn from various sites across the Grid. Component
models will be on their own grid mosaics; the Gridspec will be used to
generate custom coupler and regridding code on the basis of the
PRISM/OASIS coupler using the BFG \citep{ref:dahl1982}.

The implementation was done completely in XML. To quote the
\href{http://source.ggy.bris.ac.uk/wiki/GENIE_Gridspec}{GENIE
  Gridspec},

\begin{quote}
  The gridspec has been implemented as an XML schema in preference to
  NetCDF to fit in with the XML metadata implementation used by BFG;
  eventually the gridspec should be available in both NetCDF/CF and
  XML formats, making it accessible to a wide range of Earth system
  modelling tools and programs.
\end{quote}

Indeed, the second implementation cited here uses the netCDF-3
specification of the Gridspec.

\subsection{The GFDL implementation}
\label{sec:GridStd:GFDL}

The GFDL Earth system models have long used the \emph{exchange grid}
\citep{ref:balajietal2006} as a means of flexible transfer between
model components on independent grids. The exchange grid can be
expensive to compute, and so has always been pre-computed and stored
as a netCDF file within GFDL. As we expand the scope of our models to
include mosaics (for instance, a cubed-sphere atmospheric model), it
has become necessary to revise the grid specification. It was in the
process of this revision that the Mosaic Gridspec was devised.

The Mosaic Gridspec 0.2 specification is currently being deployed in
GFDL production codes that couple an atmosphere on a
\texttt{cube_sphere_grid}, an ocean and a sea-ice model on a
\texttt{tripolar_grid}, and a land model on a
\texttt{lat_lon_grid}. The same Gridspec is also used for
transformations of saved data between the various grids.

A complete suite of netCDF files expressing this gridspec, and a set
of C programs for generating these, are being made available through
Balaji's grid page.

\section{Summary}
\label{sec:GridStd:Summary}

The grid specification proposed serves two purposes: in various
contexts, these purposes have been described as \emph{descriptive} and
\emph{prescriptive}; \emph{semantic} and \emph{syntactic}; or
\emph{discovery} and \emph{use} metadata. The first purpose serves up
information for human consumption: attaching this metadata to model
output will enable a user to ask several key questions to understand
its content, find out whether indeed the dataset meets a given
scientific need. The second purpose is to delve further and perform
operations upon datasets: as stated in \secref{Intro:Rationale}, these
include commonly performed scientific analysis and visualization:
differential and integral calculus on vector and scalar fields; and
transformations from one grid to another. The intent is that given the
existence of a standard representation of grids, many of these
operations will be abstracted into commonly available tools and
analysis packages, and in fact may be available as web services.

An abstract representation (UML diagram) of the first class of
metadata is shown in \figref{gridspecDetails}. It is expected that
this content will eventually appear as part of a standard XML schema
to be applied to data discovery. The content of this schema will be
part of extensible controlled vocabularies to be defined by the
appropriate domain specialists.

\begin{figure}[htb]
  \centering
  \includegraphics*[width=152mm]{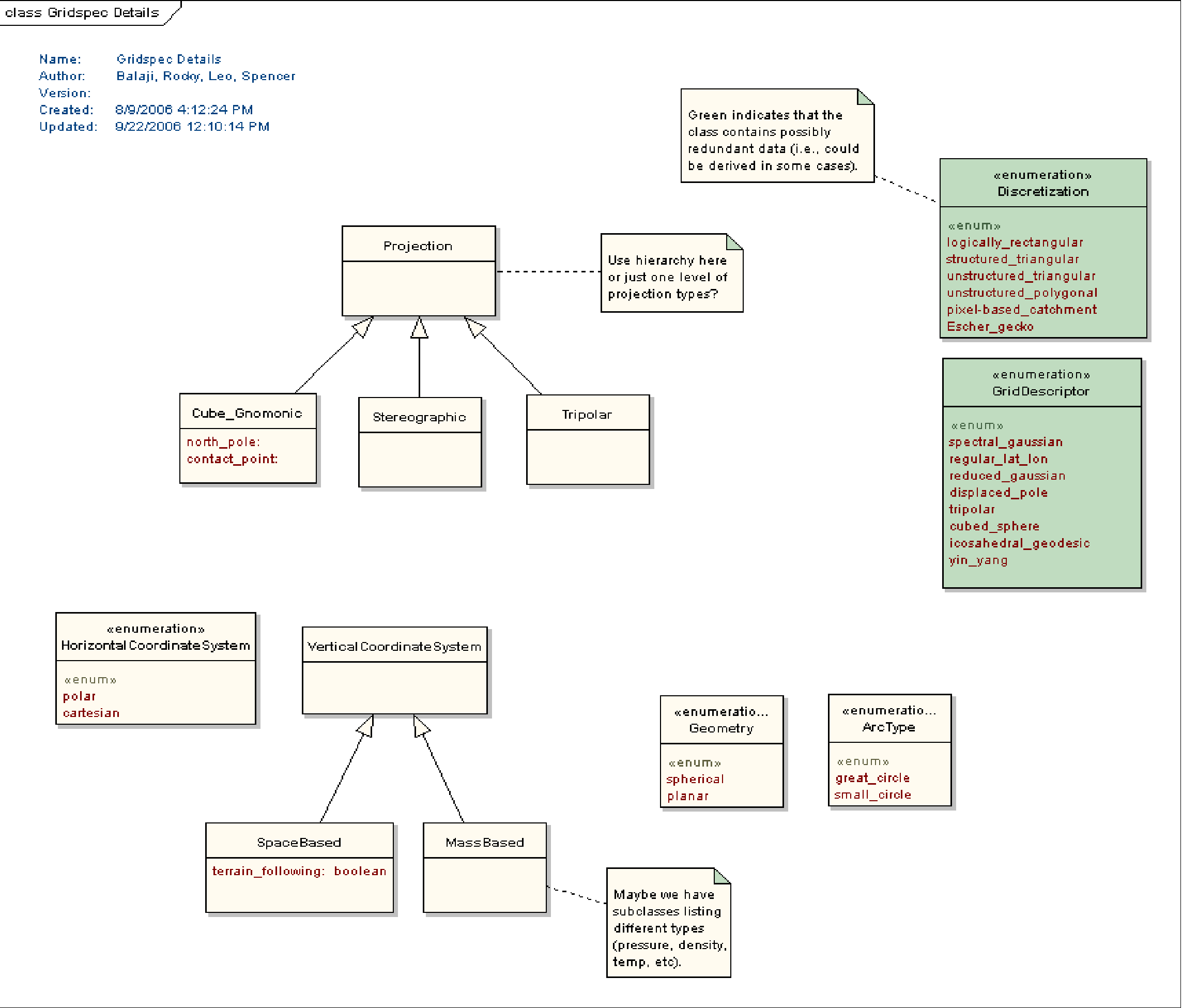}
  \caption{Semantics of grid specification. This UML diagram shows a
    vocabulary for describing grid coordinate systems, projections and
    discretizations. The proposal is that this semantic content uses a
    controlled, yet extensible, vocabulary maintained by the CF
    convention committees.}
  \label{fig:GridStd:gridspecDetails}
\end{figure}

The second class of metadata is far more detailed
(\figref{gridspecModel}). This UML diagram shows schematically how
locations, distances, areas and volumes of grid cells are conceptually
linked into a structure culminating in a grid mosaic. While also in
principle represented by a schema, these metadata are likely to be
large in size and stored in datasets in some standard data format,
netCDF being the canonical example shown here.

\begin{figure}[htb]
  \centering
  \includegraphics*[width=152mm]{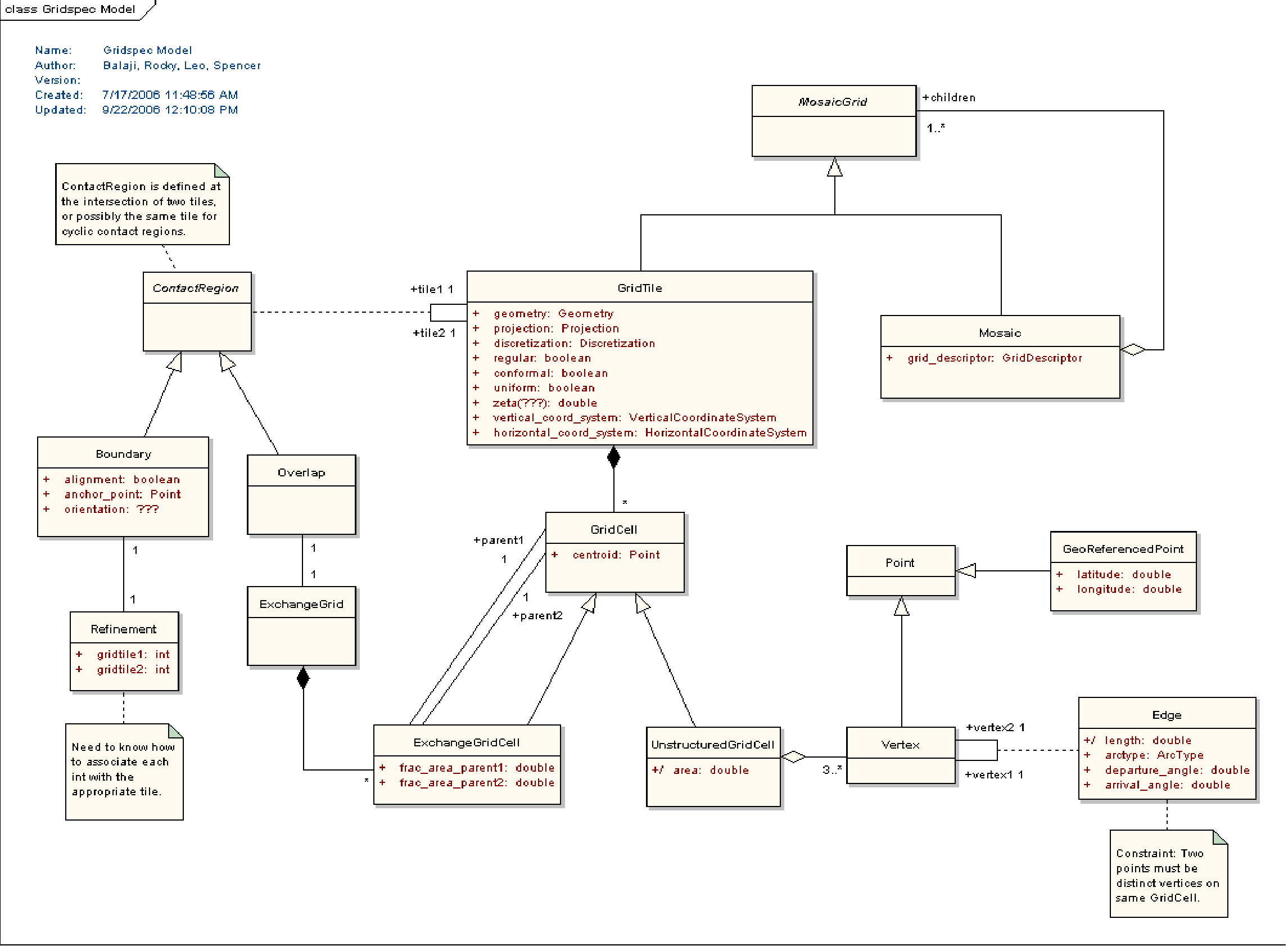}
  \caption{Abstraction of the grid specification. This UML diagram
  shows schematically how locations, distances, areas and volumes of
  grid cells are conceptually linked into a structure culminating in a
  grid mosaic.}
  \label{fig:GridStd:gridspecModel}
\end{figure}

It is possible to deduce the semantic content from the syntactic: for
instance, one could work out whether a model used a C-grid by
comparing vector and scalar field locations. Nonetheless, it would be
recommended and probably mandatory to include the very useful semantic
descriptors. Validators could be used to address the consistency
problem and ensure that the redundant information was indeed correct.

The draft specification, accompanied by prototype tools for producing
and using some example gridspec files, was released to the CF
community in early 2007.

\section*{Acknowledgments}

VB and AA are supported by the Cooperative Institute for Modeling the
Earth System, Princeton University, under Award NA18OAR4320123 from
the National Oceanic and Atmospheric Administration, U.S. Department
of Commerce. The statements, findings, conclusions, and
recommendations are those of the authors and do not necessarily
reflect the views of Princeton University, the National Oceanic and
Atmospheric Administration, or the U.S. Department of Commerce. VB
additionally acknowledges funding from the French state aid managed by
the Agence National de Recherche under the ``Investissements
d'avenir'' program with the reference ANR-17-MPGA-0010.

\label{sec:GridStd:ack}

\bibliography{refs}

\end{document}